\documentclass[sigconf]{acmart}

\AtBeginDocument{%
  }

\setcopyright{acmlicensed}
\copyrightyear{2024}
\acmYear{2024}
\acmDOI{XXXXXXX.XXXXXXX}

\acmConference[Conference acronym 'XX]{Make sure to enter the correct
  conference title from your rights confirmation emai}{June 03--05,
  2024}{Woodstock, NY}
\acmISBN{978-1-4503-XXXX-X/18/06}

\usepackage{graphicx}
\usepackage[ruled,vlined,linesnumbered]{algorithm2e}
\usepackage{makecell}
\usepackage{threeparttable}
\usepackage{makecell}
\usepackage{url}
\usepackage{soul}
\usepackage{wrapfig}
\usepackage{amsmath,amsfonts}
\usepackage{relsize}
\usepackage{tcolorbox}
\usepackage{wrapfig}
\usepackage{tikz}
\usepackage{multirow}
\usepackage{mathtools}
\usepackage{subfigure}
\usepackage{enumitem}
\usepackage{etoolbox}
\usepackage[justification=centering, font=footnotesize,labelsep=period]{caption}
\makeatletter
\patchcmd{\@makecaption}
  {\scshape}
  {}
  {}
  {}
\makeatletter
\patchcmd{\@makecaption}
  {\\}
  {.\ }
  {}
  {}
\makeatother

\definecolor{darkseagreen}{RGB}{213,232,212}
\usepackage{tikz}
\newcommand\encircle[1]{%
  \tikz[baseline=(X.base)] 
    \node (X) [draw, shape=circle, inner sep=-1.5, fill=darkseagreen] {\strut #1};}

\newcommand{\e}{{\sf eid}}
\newcommand{\id}{{\sf id}}

\begin{document}

\title{When GDD meets GNN: A Knowledge-driven Neural Connection for Effective Entity Resolution in Property Graphs}

\author{Junwei Hu$^1$, Michael Bewong$^{2,3}$, Selasi Kwashie$^3$, Yidi Zhang$^1$, Vincent Nofong$^{4}$, John Wondoh$^2$, Zaiwen Feng$^1$*}\thanks{*Correspondence: Zaiwen.Feng@mail.hzau.edu.cn}
\affiliation{%
  \institution{$^1$College of informatics, Huazhong Agricultural University, Wuhan, Hubei, China}
  \institution{$^2$School of Computing, Mathematics \& Engineering, Charles Sturt University, Wagga Wagga, NSW, Australia}
  \institution{$^3$AI \& Cyber Futures Institute, Charles Sturt University, Bathurst, NSW, Australia}
  \institution{$^4$Faculty of Engineering, University of Mines and Technology, Ghana}
}

\renewcommand{\shortauthors}{Hu et al.}

\begin{abstract}
This paper studies the entity resolution (ER) problem in property graphs. ER is the 
task of identifying and linking different records that refer to the same real-world entity. 
It is commonly used in data integration, data cleansing, and other applications where 
it is important to have accurate and consistent data.
In general, two predominant approaches exist in the literature: rule-based 
and learning-based methods. On the one hand, rule-based techniques are often desired 
due to their explainability and ability to encode domain knowledge. Learning-based
methods, on the other hand, are preferred due to their effectiveness in spite of their
black-box nature.
In this work, we devise a hybrid ER solution, {\sf GraphER}, that leverages the strengths of both systems 
for property graphs. In particular, we adopt {\em graph differential dependency} (GDD) for
encoding the so-called {\em record-matching rules}, and employ them to guide a graph neural network (GNN) based representation learning for the task. 
We conduct extensive empirical evaluation of our proposal on benchmark ER datasets including 17 graph datasets and 
7 relational datasets in comparison with 10 state-of-the-art (SOTA) techniques. 
The results show that our approach provides a significantly better solution to addressing ER in graph data, both quantitatively and qualitatively, while attaining highly competitive results on the benchmark relational datasets {\em w.r.t.} the SOTA solutions. 
\end{abstract}





\maketitle

\section{Introduction}
Entity resolution (ER) is the task of disambiguating data to determine multiple digital records that represent the same real-world entity such as a person, organization, place, or thing. ER is essential in data cleansing tasks, and can also be a powerful tool in determining fraudulent actors with multiple differing representations across databases. For example, a single adversarial actor may establish multiple social media accounts to propagate trolling behaviour, cyberbullying or even information warfare. Such multiple accounts can often be challenging to prevent 
\emph{e.g.} multiple email addresses can easily be created to satisfy email validation mechanisms. They can also be difficult to identify and eradicate due to absence of strict schema and constraints in the representation of such data. For example, an actor may have multiple accounts with slightly different contact information, and with addresses formatted differently, using different forms/abbreviations of names, etc. This problem is well-studied in the relational data setting~\cite{1,2,3,4,peeters2021dual,akbarian2022probing}.

However, the rise in popularity of unstructured data \emph{e.g.} social media and the ability of graph data to faithfully represent structural information amongst entities have made graph databases the mainstay for representing such data. In line with this, the problem of entity resolution in graph databases has become both pertinent, and non-trivial. In this work, we focus on finding different representations, structurally and semantically, of the same real world entity in graph databases.

\begin{figure}[t]
    \centering
    \includegraphics[width=.9\linewidth]{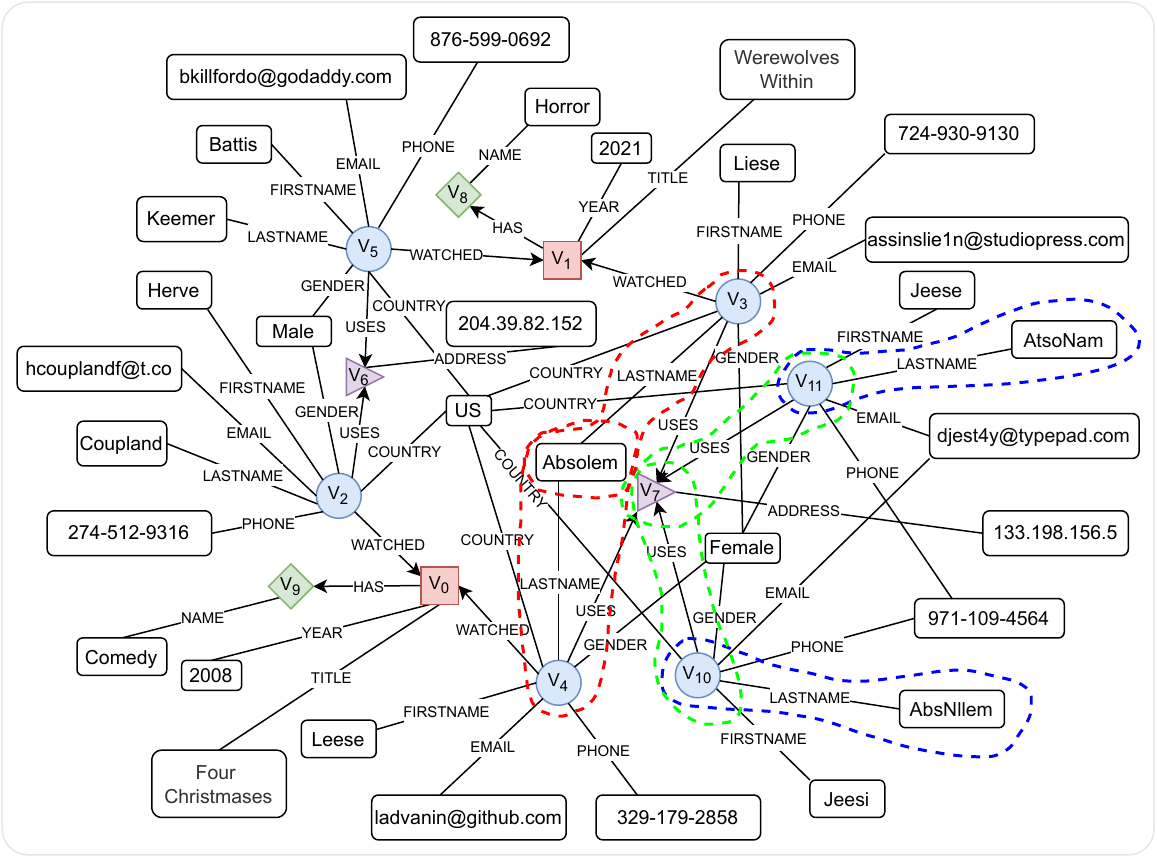}
    \vspace{-2ex}
    \caption{An exemplar property graph, $G$, of a video streaming platform}
    \label{fig:video}
    \vspace{-1.5ex}
\end{figure}

\begin{example}[Motivation]
Figure~\ref{fig:video} is a toy property graph of an online video streaming network. 
There are four types of entities in the graph $G$, namely \textbf{\emph{video}} ($v_0$, $v_1$), \textbf{\emph{user}} ($v_2$, $v_3$, $v_4$, $v_5$, $v_{10}$, $v_{11}$), \textbf{\emph{ipaddress}} ($v_6$, $v_7$), and \textbf{\emph{genre}} ($v_8$, $v_9$). 
In such a graph, a single user may create multiple {\em user} accounts to troll other users or conduct information warfare activities\footnote{e.g. Multiple YouTube accounts linked to the Russian Internet Research Agency were purportedly used in the alleged meddling of the 2016 US elections by Russia~\cite{Golovchenko2020, Howard2018ira}.}. 

Thus, for data cleansing or adversarial-actor detection tasks, it is crucial to
find all nodes in $G$ that belong to the same entity in the real-world. 
However, this turns out to be a challenging non-trivial task. Consider the 
nodes $v_3,v_4$ (in red highlights), and nodes $v_{10}, v_{11}$ (in blue highlights). Whereas the pair $v_3,v_4$ is easily identifiable as a potential
match (due to their shared LASTNAME), $v_{10}, v_{11}$ (with different LASTNAME values) 
may only appear as a candidate pair via analysis of highlighted path in green. $\square$
\end{example}

The example shows the need for a holistic approach which considers both 
structural and attribute information in graphs. 
In practice, designing ER solutions require breaking a crucial 
logjam~\cite{21}: machine learning-based solutions {\em or} rule-based solutions?
On the one hand, in spite of their relatively higher computational cost, the former are 
often favoured as they are usually more effective. Whereas on the other hand, the latter 
are desired as they are interpretable and explainable. 
In this work, we propose a holistic hybrid ER solution for property graphs to exploit the 
strengths of both worlds. 
However, three key challenges emerge: {\em effective encoding of relevant matching 
rules} (C1); {\em effective use of rules to enhance the learning models for match 
determination} (C2); and {\em equivalent accuracy and efficiency w.r.t. learning-based 
ER solutions} (C3).


The contributions of this paper are summarised as follows.
First, we propose a novel hybrid rule- and learning-driven framework 
for addressing the entity resolution problem in property graphs using both
structural and attribute information. Specifically, we employ {\em graph differential
dependencies} (GDDs)~\cite{6} to represent and encode matching rules that guide the 
representation learning over property graphs using graph neural networks (GNNs) for 
effective and explainable linking of entities. Furthermore, we present 
practical insights for discovering semantically meaningful and effective GDDs in sample
eid-labelled property graphs (C1). 
The framework demonstrates how both structural and attribute relationships 
encoded in GDDs can be used explicitly to aid the vectorization of  
an input property graph using meta-path-based and encoder-decoder graph neural networks. 
Second, we propose a four-step procedure (based on our hybrid ER framework) to 
effectively link entities in heterogeneous property graphs. The four-step solution, 
{\sf GraphER}~\footnote{All software code and datasets developed from this paper have 
been made publicly available at \url{https://github.com/Zaiwen/Entity_Resolution_Junwei_HU}}, entails: a) GDD-aided representation learning which 
imputes domain knowledge into learnt representations -- C2; b) computationally efficient 
and effective clustering of candidate nodes (\emph{a.k.a.} blocking) -- C3; c) effective 
pruning of false-positive node pairs -- C3; and d) explainable GDD-aided matching 
determination (C2).
Third, we conduct extensive experiments using established ER benchmark
datasets, viz.: 17 graph datasets and 7 relational datasets in comparison with 10  
state-of-the-art (SOTA) ER techniques. Our experiments show that, in general,
{\sf GraphER} is as effective as any SOTA deep-learning based ER solution. 
On graph datasets, {\sf GraphER} outperforms all SOTA techniques with a 95.4\% F1 score 
on average compared to 87.3\% for the rule-based graph ER method, {\sf Certus}~\cite{6};
and 81.9\% for the best learning-based model, \emph{RoBERTa}~\cite{liu2019roberta}. 
Further, the empirical results show that {\sf GraphER} is not only effective over 
graph data but also yields highly competitive results over relational datasets. 
For instance, {\sf GraphER} achieves on average 91.9\% F1 score in comparison with 
the next best relational technique \emph{RobEM}~\cite{akbarian2022probing} at 91.7\%. 
Finally, we assess the effectiveness of various components of {\sf GraphER} via a 
thorough ablation study, and show the impact of the structural and attribute
information encoded and translated by GDDs into the learning process.

\section{Preliminaries}\label{sec2}
This section covers key concepts and definitions used in the paper.


\subsection{Property Graph, Graph Pattern, \& Graph Dependency}\label{sec:gpm}
The definitions of graph, graph pattern, and matches follow those in~\cite{6,9}.
Let \textbf{A}, \textbf{L}, \textbf{C} denote universal sets of attributes, 
labels, and constants respectively.

\subsubsection{Property Graphs}
We consider a directed {\em property graph} $G = (V, E, L, F_A)$, where: (1) 
$V$ is a finite set of nodes; (2) $E$ is a finite set of edges, given by 
$E\subseteq V \times \textbf{L} \times V$, in which $(v, l, v')$ is an edge from 
node $v$ to node $v'$ with label $l\in \textbf{L}$; (3) each node $v\in V$ has 
two special attributes---$v$.\id\ denoting its identity (always present), and $v.$\e\ 
indicating the identity of the real-world entity it represents (often unknown/
unavailable)---and a label $L(v)$ drawn from $\textbf{L}$; 
(4) every node $v$, has an associated list $F_A(v) = [(A_1,c_1),\cdots,(A_n, c_n)]$ 
of attribute-value pairs, where \(c_i\) $\in$ \textbf{C} is a constant, \(A_i\) $\in$ 
\textbf{A} is an attribute of $v$, written as \(v.A_i\) = \(c_i\), and \(A_i\) $\neq$ \(A_j\) if $i \neq j$.
We say a graph, $G$, is \e-labelled if some nodes in $G$ have value
for the special attribute \e.


\subsubsection{Graph Pattern} A graph pattern, denoted by $Q$[$\bar{u}$], is a directed graph $Q$[$\bar{u}$] = ($V_Q$, $E_Q$, $L_Q$), where: (1) $V_Q$ and $E_Q$ represent the set of pattern nodes and pattern edges respectively; (2) $L_Q$ is a label function that assigns a label to each node $v$ $\in$ $V_Q$ and each edge $e$ $\in$ $E_Q$; and (3) $\bar{u}$ is all the nodes, called (pattern) variables in $V_Q$. All labels are drawn from \textbf{L}, including the wildcard “*” as a special label. 
Two labels $l$, $l'$ $\in$ \textbf{L} are said to \textbf{\em match}, denoted $l$ $\asymp$ $l'$ iff: (a) $l$ = $l'$; or (b) either $l$ or $l'$ is “*”.

\begin{figure*}[ht]
    \centering
    \includegraphics[width=.95\linewidth]{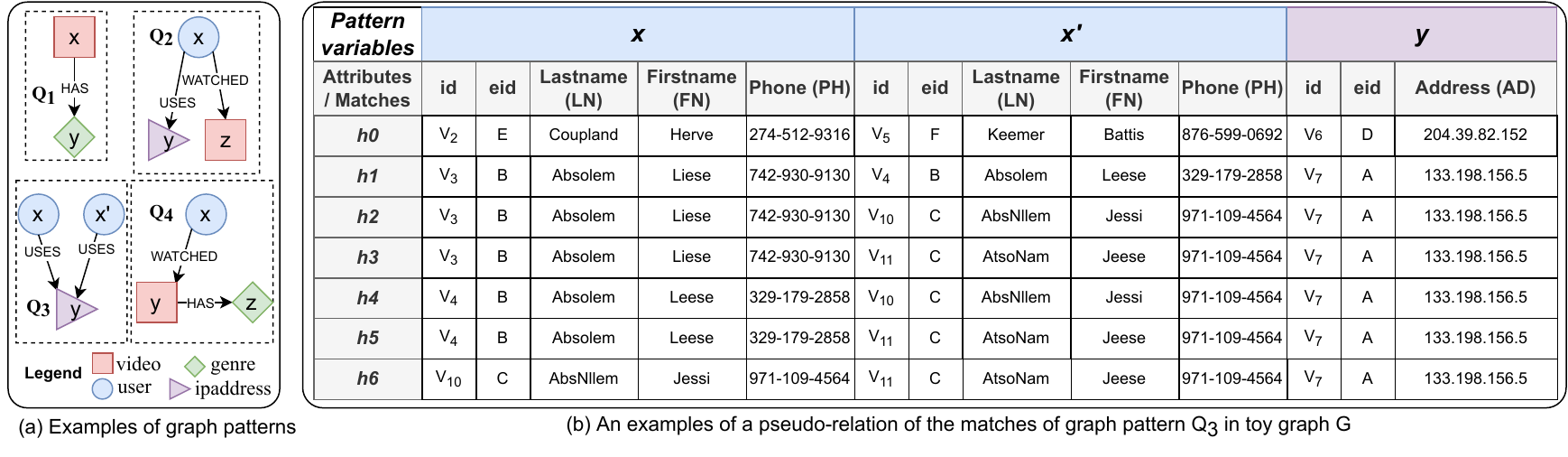}
    \vspace{-3ex}
    \caption{Graph patterns (left); and pseudo-relation of matches of pattern $Q_3$ in $G$ (right)}
    \label{fig:pattern}
    \vspace{-1.5ex}
\end{figure*}

\subsubsection{Matching graph patterns}
A \textbf{\em match} of a graph pattern $Q$[$\bar{u}$] in a graph $G$ is a homomorphism \textit{h} from $Q$ to $G$ such that: (a) for each node $v$ $\in$ $V_Q$, $L_Q$($v$) $\asymp$ $L$(\textit{h}($v$)); 
and (b) each edge $e$ = ($v$, $l$, $v'$) $\in$ $E_Q$, there exists an edge $e'$ = (\textit{h}($v$), $l'$, \textit{h}($v'$)) in $G$, such that $l$ $\asymp$ $l'$. We denote the list of all matches of $Q$[$\bar{u}$] in $G$ by \textit{H}($\bar{u}$). An example of graph patterns and their matches are presented below in \emph{Example~\ref{ex1}}.

\begin{example}\label{ex1}
\fussy
In Figure~\ref{fig:pattern}~(a) is an example of four graph patterns appearing in the property graph of Figure~\ref{fig:video}. $Q_1[x, y]$ describes a \textbf{\emph{video}} node $x$ having a {\sc has} relationship with a \textbf{\emph{genre}} node $y$. The list of matches of this pattern in the example graph is $H_1(x, y) = [\{v_0, v_9\}, \{v_1, v_8\}]$. $Q_2[x, y, z]$ depicts a \textbf{\emph{user}} node $x$ with a {\sc uses} relation with an \textbf{\emph{ipaddress}} node $y$, and a {\sc watched} relation with a \textbf{\emph{video}} node $z$. Its match in $G$ is $H_2(x, y, z) = [\{v_2, v_6, v_0\},\{v_3,v_7,v_1\},\{v_4,v_7, \\ v_0\},\{v_5,v_6,v_1\}]$. $Q_3[x,x', y]$ specifies two \textbf{\emph{user}} nodes $x$ and $x'$ each having a {\sc uses} relation with an \textbf{\emph{ipaddress}} node $y$. Thus, matches in $G$ are $H_3(x, x', y) = [\{v_2, v_5, v_6\}, \{v_3, v_4, v_7\}, \{v_3, v_{11}, v_7\}, \{v_3, v_{10}, v_7\}, \\ \{v_4, v_{10}, v_7\}, \{v_4, v_{11}, v_7\}, \{v_{10}, v_{11}, v_7\}]$. $Q_4[x, y, z]$ describes a \textbf{\emph{user}} node $x$ having a {\sc watched} relation with a \textbf{\emph{video}} node $y$, and the \textbf{\emph{video}} node $y$ also contains a {\sc has} relation with a \textbf{\emph{genre}} node $z$. $H_4(x, y, z) = [\{v_2, v_0, v_9\}, \{v_4, v_0, v_9\}, \{v_5, v_1, v_8\}, \{v_3, v_1, v_8\}]$ are its matches. $\square$
\end{example}

\subsubsection{Graph Differential Dependency~\cite{6}}
A graph differential dependency(GDD) $\varphi$ is a pair ($Q$[$\bar{u}$], ${\rm \Phi}_X$ → ${\rm \Phi}_Y$), where $Q$[$\bar{u}$] is a graph pattern, and ${\rm \Phi}_X$, ${\rm \Phi}_Y$ are two (possibly empty) sets of distance constraints on the pattern variables $\bar{u}$. A distance constraint in ${\rm \Phi}_X$ and ${\rm \Phi}_Y$ on $\bar{u}$ is one of the following:

\begin{tabular}{c|c}
   \hline
   constant-constraints (CC) & variable-constraints (VC) \\
  \hline
  $\delta_{A}(x.A, {c}) \leq {t}_{{A}};$ & $\delta_{{A}_{1}{A}_{2}}\left(x.{A}_{1}, x^{\prime}.{A}_{2}\right) \leq {t}_{{A}_{1}{A}_{2}};$ \\
  $\delta_{\e}\left(x.\e, c_{e}\right)=0;$ & $\delta_{\e}\left(x.\e, x^{\prime}.\e\right)=0;$ \\
 $\delta_{\equiv}\left(x.rela, c_{r}\right)=0$; & $\delta_{\equiv}\left(x.rela, x^{\prime}.rela\right)=0$;\\
 \hline
\end{tabular}

\noindent where $x$, $x'$ $\in$ $\bar{u}$, are pattern variables, $A,A_1,A_2$ $\in$ \textbf{A}, are non-\id\ attributes, and $c$ $\in$ \textbf{C} is a constant. $\delta_{A_iA_j}$($x.A_i$, $x'.A_j$) (or $\delta_{A_i}$($x$, $x'$) if $A_i$ = $A_j$) is a user-specified distance function for the value of ($A_i$, $A_j$), $t_{A_1A_2}$ is a threshold for ($A_1$, $A_2$), $\delta_\e$($\cdot$, $\cdot$) (\emph{resp.} $\delta_\equiv$($\cdot$, $\cdot$)) is a function on \e\ (\emph{resp.} relations) and returns 0 or 1. $\delta_\e$($x$, $c_e$) = 0 if the \e-value of $x$ is $c_e$, $\delta_\e$($x$, $x'$) = 0 if both $x$ and $x'$ have the \e-value, $\delta_\equiv$($x.rela$, $c_r$) = 0 if $x$ has a relation named {\sf rela} and ended with the node $c_r$, $\delta_\equiv$($x.rela$, $x'.rela$) = 0 if both $x$ and $x'$ have the relation named {\sf rela} and ended with the same node.
In general, constant constraints (CC) evaluate the difference between 
the value of a pattern variable (e.g., attribute-value of a node) and a specified 
constant value, whereas variable constraints (VC) compare values of two pattern variables 
(e.g., attribute-values of two nodes).

$Q$[$\bar{u}$], and ${\rm \Phi}_X$ → ${\rm \Phi}_Y$ are referred to as the pattern/scope and dependency of $\varphi$ respectively. We call ${\rm \Phi}_X$ and ${\rm \Phi}_Y$ the LHS and the RHS functions of the dependency respectively.

The user-specified distance function $\delta_{A_1A_2}$($x.A_1$, $x'.A_2$) is dependent on the types of $A_1$ and $A_2$. It can be an arithmetic operation of interval values, an edit distance of string values or the distance of two categorical values in a taxonomy, etc. The functions handle the wildcard value “*” for any domain by returning the 0 distance.


\begin{example}\label{ex2}
We illustrate the semantics of $\rm GDD_L$ via the property graph in Figure~\ref{fig:video} and the graph pattern $Q_3$ in Figure~\ref{fig:pattern}~(a).
Let a $\rm GDD_L$ $\varphi : (Q_3[x, x', y], \{\delta_{\sf FN}(x, x')\leq 0.24 \wedge \delta_{\sf LN}(x, x')=0\} \to \delta_\e(x, x')=0)$. \emph{i.e.} 
for any match of two users $x,x'$, in $G$ that {\sc uses} the same \textbf{\em ipaddress} $y$, if
the distance between the first- and last-name (of $x,x'$) as measured by the functions 
$\delta_{\sf FN},\delta_{\sf LN}$ are within \(0.24\) and \(0\) respectively, then users $x,x'$
are the same person. 
$\square$
\end{example}

\subsection{Problem Definition}\label{PD}
\begin{definition}[Entity Resolution (ER)]
Given a property graph, $G$, we seek to identify and link all node pairs
$v_i,v_j\in G$, that refer to the same real-world entity. \qed
\end{definition}

In the ER task, our method takes a property graph $G$ containing duplicate 
entities as input. Firstly, a set of candidate pairs $C_b$ is generated 
through the blocking, and then further pruned to obtain a purer set of 
candidate pairs $C_p$. Finally, matching pairs $C_m$ are generated through 
a matching step. In the following, we present how GDDs can be used to 
guide graph neural networks to achieve the aforementioned steps.

\begin{figure*}[!t]
\centering
\includegraphics[width=.7\linewidth]{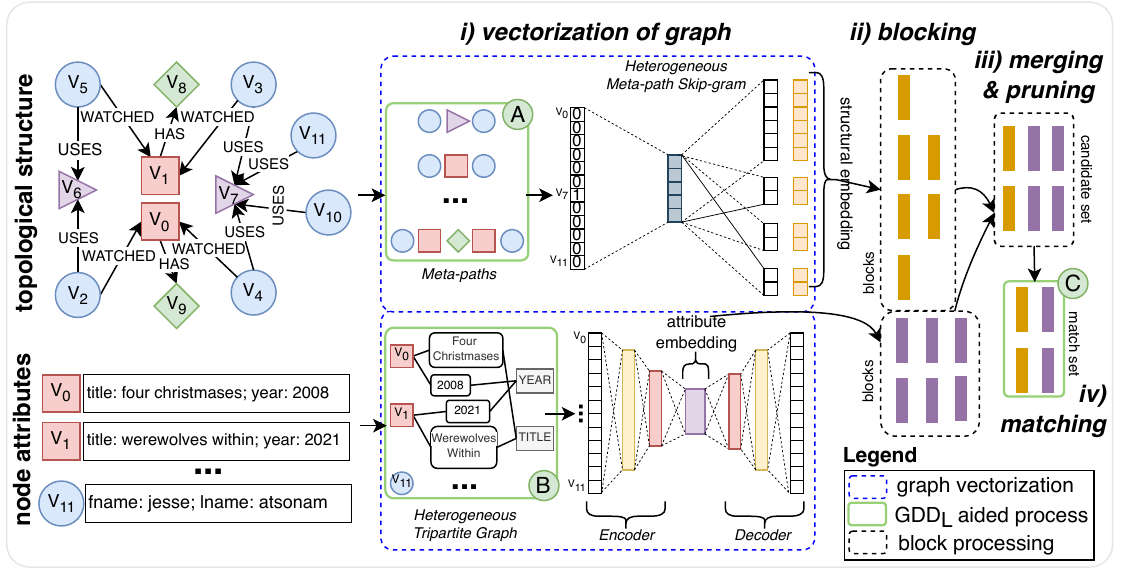}
\vspace{-2ex}
\caption{Proposed ER framework}\label{fig:pipeline}
\vspace{-2ex}
\end{figure*}

\section{The Proposed ER Framework}\label{sec3}

\subsection{Mining Graph Differential Dependencies}\label{M_GDD}
The existence of semantically meaningful entity linking rules is crucial 
to the solution. In the following, we briefly discuss how one might obtain 
such rules, encoded via GDDs for the ER task. 

We are interested in a special class of
GDDs \emph{i.e.} $\rm GDD_Ls$ where $\Phi_Y$ consists only of constraints on the \e\ of pattern variables $\bar{u}$, i.e., $\Phi_Y\ni \delta_{\e}\left(x.\e, c_{e}\right)=0 \mbox{ or } 
\delta_{\e}\left(x.\e, x^{\prime}.\e\right)=0$.

Given, a property graph, $G$, the discovery problem of $\rm GDD_Ls$ is to mine a
non-redundant set of $\rm GDD_Ls$ valid in $G$. This discovery relies on the availability of
sample eid-labelled graph. 
Figure~\ref{fig:pattern}~(b) shows matches of the graph pattern $Q_3$ in Figure~\ref{fig:pattern}~(a) in the toy graph of Figure~\ref{fig:video} as a 
pseudo-relation with attribute information (including \e-values).

In this work, we adopt 
existing GDD discovery techniques~\cite{6,zhang2023discovering}
to mine the linking rules over sample eid-labelled graph via a two-phase approach.
%
First, we find frequent graph patterns in $G$ using \verb|GraMi|~\cite{grami}
to represent the scopes (or loose-schema).
It is noteworthy that domain-experts can also specify semantically meaningful 
graph patterns to be used as the scopes of the dependency.
For each graph pattern $Q[\bar{u}]$, we find all (homomorphic) matches, $H(Q[\bar{u}])$
in $G$ using the efficient worst-case optimal join based algorithm
in~\cite{wcoj}. This is similar to the graph pattern mining and matching 
approach in~\cite{10}. 
For example, Figure~\ref{fig:pattern}~(a) shows four frequent patterns 
mined from the property graph in Figure~\ref{fig:video}.

Second, we define various distance functions over attributes of nodes in
the graph, and model the search space of candidate $\rm GDD_Ls$ using the lattice
data structure; and employ the level-wise search strategy in~\cite{6} to 
find a non-redundant set of $\rm GDD_Ls$ in $G$. 
We represent the matches of a given graph pattern as a pseudo-relation to facilitate 
the $\rm GDD_Ls$ discovery. 
For example, Figure~\ref{fig:pattern}~(b) represents a pseudo-relation 
of matches of the pattern $Q_3$ in Figure~\ref{fig:pattern}~(a). 
We refer interested readers to~\cite{6,zhang2023discovering} for details of 
the adopted GDD discovery algorithms.

We remark that, even when rule discovery algorithms exist, many rule synthesis problems
still suffers from two key challenges \emph{w.r.t.} finding semantically meaningful rules. 
First, the exhaustively large search space of candidate rules (DC1).
Second, the problem of determining effective and interpretable rules (DC2). 
In the case of GDDs, it is
impossible to alleviate DC1 without relevant domain knowledge of the data and 
heuristics. This is because, the choice of distance functions and thresholds 
over attributes are data-dependent, and crucial to the space of candidate 
GDDs. Thus, we select the most relevant distance function over attributes based on 
domain knowledge of the data, and restrict the number and range of possible thresholds
in accordance of the semantics they impute. For instance, for 
a \texttt{Phone} attribute, any sequential character-based distance function with a 
zero distance threshold is meaningful, whereas a \texttt{Name} attribute may require
various string and phonetic distance function with semantically meaningful thresholds greater
than zero. Secondly, to address DC2, given that all valid GDDs hold with a confidence of
100\% by definition, we use the support metric and the size of the antecedent function 
$|\Phi_X|$ as surrogates to rank effectiveness and interpretability of discovered rules 
respectively. In general, the higher the support of a GDD, the more persistent it is in the 
data, hence potentially more effective; and the smaller the value of $|\Phi_X|$, the more 
succinct and interpretable the rule.

\subsection{An Overview of the ER Framework}
In this work, we propose four key steps for the ER task, namely: a) vectorization of input graph (\emph{a.k.a.} embedding), b) clustering
of potential matching pairs (\emph{a.k.a.} blocking), c) elimination of non-matching 
pairs in clusters (\emph{a.k.a.} pruning), and d) determination of the matching
pairs (\emph{a.k.a.} matching).
Figure~\ref{fig:pipeline} presents an overview of our framework; the 
components 
\encircle{A},\encircle{B}, 
and \encircle{C} 
are $\rm GDD_L$ aided.

A simplified pseudo-code of our ER algorithm, {\sf GraphER}, is presented in Algorithm~\ref{alg:one}. 
Line 2 generates multiple blocks through the blocking phase, Line 3 calculates the average weight based on all pairs that exist in blocks, and Line 4 automatically learns a similarity threshold that minimizes loss recall.
Line 7-9 computes edge weights in each block graph and performs edge weight pruning.
Line 10-12 computes the dice coefficient between entity pairs and continues pruning.
Lines 13-15, detect whether the entity pairs that have not been pruned satisfy the distance constraint in 
each $\rm GDD_L$ $\varphi$ $\in$ $\Sigma$, and if so, it is stored in graph $\mathcal{G}$ 
as a linked pair. Note that $C_b$, $C_p$ and $C_m$ refer to the matching pairs after 
blocking, pruning and matching phases respectively for each block, while $\mathcal{G}$ 
stores all the matching pairs $C_m$ from all the blocks.

\begin{algorithm}[ht]
\caption{\sf GraphER}\label{alg:one}
\footnotesize
    \SetKwInOut{Input}{Input}\SetKwInOut{Output}{Output}
    \Input{A property graph $G$ without \e s, set $\Sigma$ of $\rm GDD_Ls$ \;}
    \Output{Linked entity graph $\mathcal{G}$ \;}
    \BlankLine
    $\mathcal{G}$ = $\emptyset$ \;
    multiple blocks (each block denoted by $B$) generated through blocking phase \tcp*{\textcolor{blue}{Blocks generation reference Section~\ref{sec:blocks}}}
    calculate average harmonic mean $\rm {\bf{avW}}$ from these blocks \;
    learn a dice threshold $\vartheta$, minimum loss recall \;
    \For{each block graph $G_B$ = $(V_B, E_B)$}{
        \For{each edge $e$ = $(v, v')$ $\in$ $E_B$}{
            compute $W(v, v')$ through Eq (\ref{eq1}) \;
            \If{$W(v, v')$ $\textless$ $\rm {\bf{avW}}$}{prune edge $e \in E_B$, goto Line 6 \; \tcp{\textcolor{blue}{Pruning with heuristic rule (cf. Section~\ref{sec:blocks})}}}
            compute ${\rm dice}(v, v')$ through Eq (\ref{eq2}) \;
            \If{${\rm dice}(v, v')$ $\textless$ $\vartheta$}{prune edge $e \in E_B$, goto Line 6 \; \tcp{\textcolor{blue}{Pruning with similarity (cf. Section~\ref{sec:blocks})}}}
            \For{each $\varphi$ $\in$ $\Sigma$}{
                \If{entity pair $(v, v')$ satisfy all distance constraints from $\varphi$}{add linked pair ($v$, $v'$) to $\mathcal{G}$ \; \tcp{\textcolor{blue}{Matching determination reference Section~\ref{sec:match}}}
        }}}}
    \Return{$\mathcal{G}$} \;
\end{algorithm}

\section{Vectorization of input graph}\label{sec:g2vec}
In this section, we discuss the vectorization of the input property
graph, $G=(V,E,L,F_A)$. The goal here is to learn a latent representation, 
$\mathcal{F} \in \mathbb{R}^{|V|\times d}, d\ll|V|$, of $G$ that preserves:
a) the proximity between nodes and their neighbourhoods, as well as 
b) node properties and relations amongst properties. 

\subsection{Learning Structural Information}\label{sec:learning structure}
Here, we attempt to embed the closeness relationship between nodes and
their neighbours using semantic \textit{meta-paths} (\emph{cf.}~\cite{metapath2vec})
generated from graph patterns of the linking GDDs. 

A meta-path, $\mathtt{mp}$, is defined as a path in 
a graph of the form: $L_1 \xrightarrow{l_1}  L_2\xrightarrow{l_2} \cdots 
 \xrightarrow{l_{t-1}} L_t$ (abbreviated as $L_1$-$L_2$- $\cdots$ -$L_t$); 
 describing a composite relation $l = l_1 \circ l_2 \circ l_{t-1}$ between node types 
 $L_1$ and $L_t$, where $\circ$ represents the composition operator on relations.

\subsubsection{Meta-path-based Skip-Gram}
The idea of node-neighbourhood preservation in a graph is analogous to word-context 
preservation in corpus as seen in the \verb|word2vec|~\cite{word2vec} model.
Intuitively, given a node $v\in G$, we want to maximise the probability of 
having the neighbours, $N(v)$, of node $v$.
More formally, we use skip-gram to learn the structural representation of a 
node $v\in G$, by maximising the following probability~\cite{metapath2vec}:
\begin{align}\label{eq:obj}
    \mathlarger{arg\max_{\theta} \sum_{v\in V} \sum_{l\in L(V)} \sum_{u_l\in 
    N_l(v)} \log p(u_l|v;\theta)}
\end{align}
where $l$ is a label in the set of node labels $L(V)$, and
$N_l(v)$ is the neighbours of the node $v$ of type (or with label) $l$.
The probability $p(u_l|v;\theta)$ is modeled with a softmax function as 
$\frac{\exp(\mathcal{F}(u_l)\cdot {\mathcal F}(v))}{\sum_{w\in V}
\exp({\mathcal F}(w)\cdot {\mathcal F}(v))}$, where ${\mathcal F}(\cdot)$ is the 
$d$-dimensional vector representation of a node.

However, in practice, we use the negative sampling optimisation~\cite{word2vec} to 
construct the softmax using $k$-sized samples. That is, $p(u_l|v;\theta)$ 
in Equation~\ref{eq:obj} is approximated via softmax function as:
\begin{align}
        \sigma\left({\mathcal F}(u_l)\cdot {\mathcal F}(v) + \sum_{i=1}^{k} \mathbb{E}_{w^{(i)}
    \sim {\mathcal P}(w)}\left[\log \sigma(-{\mathcal F}(w^{(i)})\cdot {\mathcal F}(v)\right]\right)
\end{align}
where $\sigma(y) = \frac{1}{1+\exp(-y)}$, and ${\mathcal P}(w)$ is the defined 
negative sampling distribution (e.g., a custom unigram distribution
). 

\subsubsection{GDD-scopes as Meta-paths}
Given a set $\Sigma$, of linking GDDs; let  ${\mathcal Q}$ denote the set of all 
graph patterns in $\Sigma$ \emph{i.e.} ${\mathcal Q}=\{Q[\bar{u}]\mid  \exists\   
(Q[\bar{u}], \Phi_X\to \Phi_Y)\in \Sigma\}$. 
For each $Q[\bar{u}]\in {\mathcal Q}$, we generate symmetric meta-path schemes 
\verb|mps|$(Q[\bar{u}])$, based on $Q[\bar{u}]$ as follows:
\begin{align}
    \smaller{
    \mathtt{mps}(Q[\bar{u}]) = \{ mp\},
    }
\end{align}
where (1) $mp \coloneqq x_1 \xrightarrow{l} \cdots 
    \to x_t\xrightarrow{l'} x_{t+1} \cdots \xrightarrow{l} x_1$; (2) $x\in \bar{u}$; and (3) for any two node sequences $x_i, x_j\in mp$, there exists 
an edge between $x_i,x_j$ in the graph pattern $Q[\bar{u}]$. 
Each meta-path $mp\in \mathtt{mps}$ is used to guide the random walk over
$G = (V, E, L, F_A)$ (\emph{cf.} component \encircle{A} in Figure~\ref{fig:pipeline}).

Given a meta-path $mp\coloneqq x_1 \xrightarrow{l} \cdots \to x_t\xrightarrow{l'} x_{t+1} \cdots 
\xrightarrow{l} x_1$, the transitional probability at step $i$ in the random walk over $G$ (e.g., 
corresponding to node variable type $x_t$ in the meta-path) is:
\begin{align}\label{eq:prob}
    {
        p(v^{(i+1)}|v_t^{(i)}, mp) = 
            \begin{cases}
                \frac{1}{|N_{t+1}(v_t^{(i)})|}, & (v^{(i+1)}, v_t^{(i)})\in E,\\&L(v^{(i+1)})\asymp L(x_{t+1})\\\\
                0, 
                & (v^{(i+1)}, v_t^{(i)})\in E,\\ & L(v^{(i+1)})\not\asymp L(x_{t+1})\\\\
                0, & (v^{(i+1)}, v_t^{(i)})\notin E
            \end{cases}
            }
\end{align}
where $v_t^{(i)}\in V$ is the node $v$ of type $x_t$ in the meta-path at the $i^{th}$ step in the random walk; $N_{t+1}(v_t^{(i)})$ is a set of $t+1$ labelled neighbouring
nodes of node $v_t^{(i)}$ \emph{i.e.} $t+1$ indicates the next node type $x_{t+1}$ in the meta-path sequence $mp$ after node type $x_{t}$; and $L(v^{(i+1)})\asymp L(x_{t+1})$ means that the label of next node $v^{(i+1)}$ and next node type $x_{t+1}$ match. Note that, unlike in~\cite{metapath2vec}, we adopt a relaxed label matching semantics
(as seen in Section~\ref{sec:gpm}).

\begin{example}[Graph patterns as meta-paths]\label{ex3}
Consider the four entity types \textbf{\emph{video}} (v), \textbf{\emph{user}} (u), \textbf{\emph{ipaddress}} (i), and \textbf{\emph{genre}} (g) in Figure~\ref{fig:video} with the aim of identifying the same user account. $Q_3$ and $Q_4$ in Figure~\ref{fig:pattern}~(a) can be captured as symmetric meta-path schemes as follows: 
\verb|mps|$(Q_3[\bar{u}])$ = \{u-i-u\} \emph{i.e.}, $(u) \xrightarrow{uses} (i)\xleftarrow{uses}
(u)$  and \verb|mps|$(Q_4[\bar{u}])$ = \{u-v-g-v-u\} \emph{i.e.} $(u)\xrightarrow{watched}
(v)\xrightarrow{has}(g)\xleftarrow{has}(v)\xleftarrow{watched}(u)$. 
The meta-path $mp$(\verb|u-i-u|) encodes the structural relation between two \verb|user| nodes 
using the same \verb|ipaddress|; while the meta-path $mp$(\verb|u-v-g-v-u|) captures 
two \verb|user| nodes' \verb|watched| structural relation with videos of the same \verb|genre|. 
The generated meta-paths guide the random walk in Figure~\ref{fig:video}. 
Following Equation~(\ref{eq:prob}), the transition probability, we have the following 
sequence generation (with a randomly selected seed node): 
$mp$ = u-i-u : \{$v_i$-$\cdots$-$v_7$-$\cdots$-$v_j$, $i,j \in [3,4,10,11]$\} and \{$v_i$-$\cdots$-$v_6$-$\cdots$-$v_j$, $i,j \in [2,5]$\},
$mp$ = u-v-g-v-u : \{$v_i$-$\cdots$-$v_0$-$v_9$-$v_0$-$\cdots$-$v_j$, $i,j \in [2,4]$\} and \{$v_i$-$\cdots$-$v_1$-$v_8$-$v_1$-$\cdots$-$v_j$, $i,j \in [3,5]$\}.
These sequences are used as token-context inputs for the meta-path-based Skip-Gram for 
the structural embedding. \qed
\end{example}

\subsection{Learning Attribute Information}\label{sec:learning attribute}
Given a property graph $G=(V,E,L,F_A)$, each node $v\in V$ has an associated
list of attribute-value pairs $F_A(v) = [(A_1,c_1), ...,(A_n, c_n)]$. 
The second vectorization of $G$ in our framework entails the encoding 
of node features based on the canonical set, $A_\Sigma$, of attributes in the 
set $\Sigma$ of linking GDDs. 

\subsubsection{Attributes Auto-encoder} 
We use an encoder-decoder feed-forward neural network to learn the latent
representation of node properties and relationships among the properties. 
Our auto-encoder architecture consists of: an aggregator, an encoder, and 
a decoder. 

\textbf{The Aggregator} can be seen as a preprocessor for the auto-encoder -- its
goal is to learn the distributed representations of attribute-values on a node
and aggregate them as input to the encoder. 
We adapt the embedding approach in~\cite{embedi} to create a heterogeneous 
object/node-token-attribute (tripartite) graph for a guided random walk for learning 
the distributed representation of attribute values. 
The tripartite graph is shown as component~\encircle{B} in Figure~\ref{fig:pipeline}.
It derives its attribute set from the $A_\Sigma\ni X,Y$ of all attributes that appear 
in the mined GDDs ($Q[\bar{u}], \Phi_X\to \Phi_Y\in \Sigma$). 
Thus, the representation of each attribute-value/token is learnt via a
random walk over the tripartite graph with implicit attribute relations in $A_\Sigma$.
We then aggregate the list of attribute values over a node using the SIF~\cite{sif} 
algorithm, which computes the weighted average of each embedding to obtain an aggregate vector, to achieve node-level attribute embedding. 
All node-level embeddings are then fed into the encoder as input for the resultant
node-level attribute embedding.

\textbf{The Encoder-Decoder} architectures consist of simple two layer feed-forward 
neural networks (NNs). The encoder $E$ takes the input node-level embedding 
$\mathbf{i}$ (from the aggregator) and returns an output $\mathbf{l}$ which is then 
fed into the decoder $D$ to return an output $\mathbf{o}$. The objective function is 
to obtain $\mathbf{o}$ which approximates $\mathbf{i}$. Formally, this is a minimisation problem of the training loss function $\|\mathbf{o}-\mathbf{i}\|_2^2$, the squared 
$l_2$ distance. Thus the output $\mathbf{l}$ of the encoder $E$ becomes the 
embedding that represents the node-attributes. 

At the end of the vectorization of the input graph, each node will have 
two embeddings: a structural and an attribute embedding. 

\begin{figure*}[!t]
    \centering
    \includegraphics[width=0.98\linewidth]{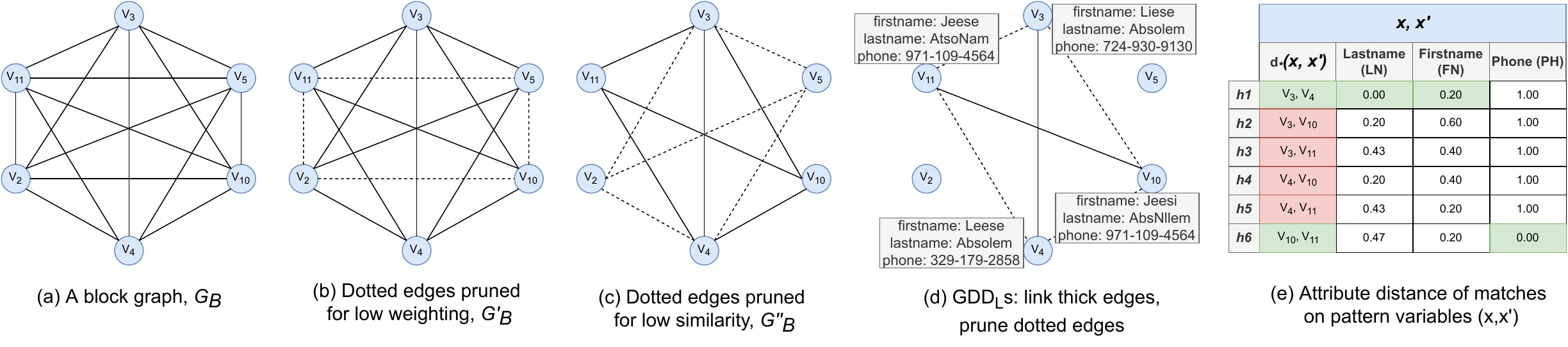}
    \vspace{-2ex}
    \caption{Blocks generation, pruning, and matching}
    \label{fig:Runing Example}
\end{figure*}

\section{Blocks generation and pruning}\label{sec:blocks}
\emph{Blocking phase.} generates a candidate set of matches. In the blocking phase, 
Locality Sensitive Hashing (LSH) is applied to the structural and the attribute vector 
representations of the entities within their embedding spaces respectively. Specifically 
we adopt the FALCONN algorithm \cite{27}, which is a well-known LSH-based 
method to solve the nearest neighbor search problem in high-dimensional spaces. 
Its goal is to find entities/vectors that are within a user-specified maximum distance of 
any nearest neighbor vector from the query. We treat each entity in the property graph $G$ 
as a query and return the most similar candidates that satisfy the distance threshold. The 
set composed of the query entity and candidates is called a block. That is, 
we obtain one block formed from structural embedding space, and the other from attribute 
embedding space for each query. In order to maximize \emph{Recall}, the two blocks are merged based on the query node. Every pair in a block may form a candidate pair. Thus for any block $B$, there will be $C_b$ candidate pairs \emph{i.e.} $|C_b| = (|B| \times (|B|-1))/2$. This is represented in the form of a block graph $G_B$ illustrated
Figure~\ref{fig:Runing Example}~(a). In the figure, there are six \textbf{\emph{user}} entities, which are obtained through the meta-path guided structural embedding (\emph{cf.} \emph{Example~\ref{ex3}}). Since every pair in the block is considered a  potential match, 15 candidate pairs $C_b$ can be generated from this block. 

\noindent\emph{Pruning phase.} aims to minimize the number of false positives within the blocks.  Let $G_B$ = ($V_B$, $E_B$), be a block graph where the node set $V_B$ is the set of all entities in the block, the undirected edge set $E_B$ contains edges ($v$, $v'$) if $v$ and $v'$ cooccur in the block $B$.\\ We propose two pruning strategies based on the edge weight and the dice coefficient as follows.

\underline{\em Pruning with edge weight:} The average weight of the edges are calculated, and
edges with a weight below the average weight is pruned. 
The weight of an edge, $e$, is the harmonic mean of the two components, \textit{Arcs, Cbc}, 
proposed in\cite{17}:
\begin{equation}\label{eq1} 
   \textit{W}(e) = 2 \cdot \frac{\textit{Arcs}\cdot\textit{Cbs}}{\textit{Arcs} + \textit{Cbs}}, 
\end{equation}
where $e$ is an edge between two nodes $v, v'\in G_B$. 
$Cbs(e) = |\mathbb{B}(e)|/\beta$;
$Arcs(e) = \frac{1}{\alpha}\sum_{B_i\in \mathbb{B}(e)}\frac{1}{|B_i|}$;
where $\mathbb{B}(e)$ is the set of all blocks in which the edge $e$ occurs; and $\alpha,\beta$ 
are normalising factors. 
$Cbs$ considers the number of blocks containing an edge and the $Arcs$ 
represents the sum of the inverse of the size of all blocks containing the edge. 
Thus the more blocks an edge occurs in the higher its $Cbs$ value. The smaller the blocks containing an edge the higher the $Arcs$ value.

We calculate the weight of every edge in $G_B$ via Equation~(\ref{eq1}) and denoted 
the average as \textbf{avW}. With the calculated average \textbf{avW}, for any edge $e\in E_B$, 
if $\textit{W}(e) < \textbf{avW}$, $e$ is deleted from $E_B$. The block graph after this step 
is denoted by $G'_B$ (\emph{cf.} Figure~\ref{fig:Runing Example}~(b)).

\underline{\em Pruning with dice coefficient:} In property graph $G$, the similarity of some attribute values play an important role in ER. For example, when the attribute LASTNAME for two nodes  has the same value \textbf{Absolem} but their attribute FIRSTNAME has values \textbf{Leese} and \textbf{Liese} respectively,
it is highly possible that \textbf{Leese} and \textbf{Liese} represent 
the same entity. This pruning method computes a dice coefficient (with consideration of value-similarity) of specific attribute values for edges (entity pairs) in the block graph $G'_B$ and other 
similarity functions also can be used (\emph{e.g.} Jaccard) in this step. Consider the edge ($v$, $v'$) in $G'_B$. The dice coefficient, denoted by dice($v$, $v'$) $\in$ [0, 1], of ($v$, $v'$) is defined as:
\begin{equation}
 {\rm dice}(v, v') = 2 \cdot \frac{\vert {v_A}\cap{v'_A} \vert}{\vert {v_A} \vert + \vert {v'_A} \vert}, A \in A^\dag \label{eq2}
\end{equation}
where $A^\dag$ is the set of common attributes of $v$ and $v'$, $v_A$ and $v'_A$ is the character set of $v$ and $v'$ on attribute $A$, respectively. If $A^\dag$ $\in$ $\emptyset$, this edge ($v$, $v'$) will be directly pruned because we consider that they are different entity types.
The dice($v$, $v'$) is calculated for all edges in the reduced block graph $G'_B$. 
We can automatically learn the most appropriate dice($v$, $v'$) threshold $\vartheta$ that minimizes mismatched entity pairs without significant loss of recall. 
Any edge ($v$, $v'$) with dice($v$, $v'$) $\textless$ $\vartheta$ is pruned from $G'_B$. We denote the updated block graph after this step by $G^{''}_B$, as shown in Figure~\ref{fig:Runing Example}~(c). The matching pairs after the pruning step is denoted by $C_p$.

\section{Match Determination}\label{sec:match}
Match determination is the final step in our ER solution \emph{i.e.} determining 
whether an entity pair refer to the same real world entity (\emph{cf.}~component~\encircle{C} in Figure\ref{fig:pipeline}). The linking decision is determined 
by a set $\Sigma$ of $\rm GDD_Ls$ learned from the eid-labelled graph described 
in Section~\ref{sec3}. 
An entity pair with edge $(v, v') \in G^{''}_B$ (\emph{cf.}~Figure~\ref{fig:Runing Example}~(c)) 
is linked if it satisfies at least one $\rm GDD_L$ $\varphi\in\Sigma$. 
More accurately, a pair $(v, v')\in G^{''}_B$ is linked if and only if $v,v'\in h$ 
such that $h\models \varphi\in \Sigma$ and $h(v,v')\asymp x,x'\in\varphi.\Phi_\e$.
We demonstrate how matching confirmations are made in \emph{Example~\ref{ex4}} below.

\begin{example}[Confirmation of matches]\label{ex4}\sloppy
Let the set of matching rules, $\Sigma:=\{\varphi_1,\varphi_2\}$, where:
\begin{itemize}
    \item $\varphi_1 : (Q_3[x, x', y],\delta_{\sf PH}(x, x') = 0 \to \delta_\e(x,x')=0)$; and
    \item $\varphi_2: (Q_3[x, x', y], \{\delta_{\sf LN}(x, x')\leq 0.25 \wedge 
    \delta_{\sf FN}(x, x') \leq 0.30\} \to \delta_\e(x,x')=0)$.
\end{itemize}
Recall the pseudo-table of all matches of the graph pattern $Q_3[x,x',y]$ 
are in Figure~\ref{fig:pattern}~(b). 
Given the candidate pairs of nodes (i.e., the remaining edges) in the 
pruned blocking graph $G^{''}_B$ in Figure~\ref{fig:Runing Example}~(c), 
we make the match confirmation decision as follows. 
Note that the candidate pairs in $G^{''}_B$ are within matches of
$Q_3[x,x',y]$: $(v_3, v_4)\in h_1, (v_3, v_{10})\in h_2, (v_3, v_{11})\in h_3,
(v_4, v_{10})\in h_4, (v_4, v_{11})\in h_5, (v_{10}, v_{11})\in h_6$.

Thus, for any candidate pair to be confirmed, their match must agree on $\varphi_1$
or $\varphi_2$. Figure~\ref{fig:Runing Example}~(e) shows the distance between
the pattern variables as measured by the distance functions in $\varphi_1$ and
$\varphi_2$. 
Evaluation of the distance constraints of the dependency (\emph{w.r.t.} 
Figure~\ref{fig:Runing Example}~(e)) show that match $h_1$ and $h_6$ agree on 
$\varphi_2$ and $\varphi_1$ respectively; whereas $h_2,h_3,h_4,h_5$ violate both.
Hence, the pairs $(v_3,v_4)$ and $(v_{10},v_{11})$ are confirmed matches, thus
refer to the same real-world entities. $\square$
\end{example}

\section{Empirical Evaluation}\label{sec:exp}
In this section we conduct extensive experiments to evaluate the efficacy of our 
approach in comparison with the SOTA.

\begin{table}[t]
\caption{Relational ER Benchmark Datasets}
\vspace{-7px}
\label{Tab:reladatasets}
\resizebox{\linewidth}{!}{
\begin{tabular}{|c|c|c|c|c|c|}
\hline
Dataset & \#Nodes & \#Edges & \#Node Types & \#Edge Types & \#Nodes with duplicates\\
\hline
Fodors-Zagats (FZ)~\cite{fz} & 1685 & 1728 & 3 & 2 & 110 \\
\hline
DBLP-ACM (DA)~\cite{dblp_acm} & 8699 & 9820 & 3 & 2 & 2220 \\
\hline
DBLP-Scholar (DS)~\cite{dblp_acm} & 135874 & 133758 & 3 & 2 & 5,347 \\
\hline
Walmart-Amazon (WA)~\cite{dblp_acm} & 27355 & 49256 & 3 & 2 & 962 \\
\hline
Amazon-Google (AG)~\cite{dblp_acm} & 4980 & 4589 & 2 & 1 & 1167 \\
\hline
BeerAdvo-RateBeer (Beer)~\cite{beer} & 11397 & 7345 & 2 & 1 & 68 \\
\hline
iTunes-Amazon (iA)~\cite{beer} & 72066 & 188490 & 4 & 3 & 132 \\
\hline
\end{tabular}
}
\vspace{-1ex}
\end{table}

\begin{table}[t]
\caption{Graph ER Benchmark Datasets}
\vspace{-7.5px}
\label{Tab:graphdatasets}
\resizebox{\linewidth}{!}{
\begin{tabular}{|c|c|c|c|c|c|}
\hline
Dataset & \#Nodes & \#Edges & \#Node Types & \#Edge Types & \#Nodes with duplicates\\
\hline
ArXiv~\cite{arxiv} & 88070 & 58515 & 2 & 1 & 5924 \\
\hline
CiteSeer~\cite{arxiv} & 4393 & 2892 & 2 & 1 & 456 \\
\hline
Entity Resolution (ER)~\cite{er_data} & 1237 & 1819 & 4 & 3 & 12 \\
\hline
WWC-1 & 2688 & 16008 & 5 & 9 & 202 \\
\hline
WWC-2 & 2688 & 15907 & 5 & 9 & 202 \\
\hline
WWC-3 & 2688 & 15757 & 5 & 9 & 202 \\
\hline
WWC-4 & 2496 & 14837 & 5 & 9 & 10 \\
\hline
WWC-5 & 2506 & 14874 & 5 & 9 & 20 \\
\hline
WWC-6 & 2587 & 15361 & 5 & 9 & 101 \\
\hline
WWC-7 & 2890 & 16801 & 5 & 9 & 404 \\
\hline
GDS-1 & 8977 & 84940 & 5 & 5 & 350 \\
\hline
GDS-2 & 8977 & 80821 & 5 & 5 & 350 \\
\hline
GDS-3 & 8977 & 80365 & 5 & 5 & 350 \\
\hline
GDS-4 & 8644 & 73602 & 5 & 5 & 17 \\
\hline
GDS-5 & 8662 & 74456 & 5 & 5 & 35 \\
\hline
GDS-6 & 8802 & 78625 & 5 & 5 & 175 \\
\hline
GDS-7 & 9327 & 90946 & 5 & 5 & 700 \\
\hline
\end{tabular}
}
\vspace{-1ex}
\end{table}

\subsection{Experimental Setup}
All experiments are performed on a 2.40GHz Intel Xeon Silver 4210R 
processor computer with 32GB memory running Linux OS. 

\subsubsection{Datasets}\label{sec:datasets} 
The datasets used for the experiments comprises of well-known relational  and 
graph ER benchmark datasets. 
Table~\ref{Tab:reladatasets} summarises commonly used relational ER benchmark datasets~\cite{1,2,3,4} while Table ~\ref{Tab:graphdatasets} shows the graph ER benchmark datasets used in this paper.
Particularly, \emph{ArXiv} and \emph{CiteSeer} are well-known citation network graph ER benchmark datasets ~\cite{bhattacharya2006latent, bhattacharya2007collective}.  
\emph{Entity Resolution} is a dataset derived from a virtual online video streaming platform. 
%
The WWC-* dataset relates to the 2019 \emph{Women’s World Cup} which contains five node types representing all the \textbf{\emph{person}}, \textbf{\emph{team}}, \textbf{\emph{squad}}, \textbf{\emph{tournament}} and \textbf{\emph{match}} from all the World Cups between 1991 and 2019; there are 9 edge types. 
%
GDS-* datasets refers to the \emph{Graph Data Science} dataset which shows the connections between different airports around the world. 
There are five different node types relating to \textbf{\emph{airport}}, \textbf{\emph{city}}, \textbf{\emph{region}}, \textbf{\emph{country}} and \textbf{\emph{continent}}, and five different edge types. Seven variations each were derived for WWC-* and GDS-* as follows:

\begin{itemize}[leftmargin=*]
\item WWC-1 (\emph{resp.} GDS-1) contains 10\% duplicate person entity type nodes (\emph{resp.} airport entity type). Duplicate nodes contain attribute noise by using an error function that applies at random, the following edits: (a) attribute values remain unchanged, (b) attribute values are changed to have a distance of 2 character difference between original and changed value, (c) attribute value is deleted. 

\item WWC-2 (\emph{resp.} GDS-2) contains 10\% duplicate nodes similar to WWC-1 (GDS-1 \emph{resp.}) however, the duplicates contain structural noise instead of attribute noise. That is, duplicate entities will have, at random, up to 50\% of its edges deleted. 

\item WWC-3 (\emph{resp.} GDS-3) contains 10\% duplicates, however the duplicate nodes contain both attribute and structural noise. 

\item WWC-4 (\emph{resp.}, GDS-4), WWC-5 (\emph{resp.}, GDS-5), WWC-6 (\emph{resp.} GDS-6) and WWC-7 (\emph{resp.} GDS-7) all contain both attribute and structural noise, however the number of matching pairs are 0.5\%, 1\%, 5\% and 20\% respectively. 
\end{itemize}

Although our approach, {\sf GraphER}, is designed for graph data, 
we demonstrate that it is also effective in addressing ER tasks in 
the relational data setting via simple data transformations~\footnotemark.

\subsubsection{Baseline Techniques} 
{\sf GraphER} is evaluated \emph{w.r.t.} the following SOTA techniques.

\noindent {\textbf{LDA-ER}~\cite{bhattacharya2006latent}: This is a probability model based on Latent Dirichlet Allocation (LDA) for collective entity resolution.}

\noindent {\textbf{CR}~\cite{bhattacharya2007collective}: This is a collective entity resolution algorithm (CR) using graph patterns and similarity measures.}

\noindent \textbf{DeepMatcher (DM)}~\cite{1}: This is a hybrid ER approach based on Smooth Inverse Frequency (SIF), Recurrent Neural Networks (RNN), and Attention-based mechanisms.


\noindent \textbf{BERT}~\cite{devlin2018bert}: A pre-trained deep bidirectional language model.

\noindent \textbf{RoBERTa}~\cite{liu2019roberta}: A robustly optimized BERT.

\noindent\textbf{Ditto}~\cite{2}: This is a novel ER system based on pre-trained Transformer based language models.


\noindent \textbf{JointBERT}~\cite{peeters2021dual}: A dual-objective training method to predict the entity identifier.

\noindent \textbf{HierGAT (HG)}~\cite{3}: This relies on a hierarchical graph attention transformer network. 


\noindent \textbf{RobEM}~\cite{akbarian2022probing}: This a pre-trained language model-based entity matching method.

\noindent \textbf{Certus}~\cite{6}: This is a rule-based ER solution on graphs.

In order to apply the baseline deep-learning techniques to the graph datasets, each dataset is first converted to a relational dataset using a data transformation tool\footnotemark[\value{footnote}]\footnotetext{Data conversion tools i.e. Relational$\rightarrow$Graph and Graph$\rightarrow$Relational Data available at: \url{https://github.com/HJW577X/data-conversion}\label{code}}.

\subsubsection{Evaluation Criteria and Metrics} 
In evaluating the baseline techniques, a $60:20:20$ split was used for training, validation and testing respectively; and report 10-fold cross-validated test results. 

Following the convention in previous work \cite{1, 4, Filtering, 11, 21}, we adopt the following common evaluation metrics and propose a new one called \emph{purity}: 
\begin{itemize}[leftmargin=*]
\item \textbf{Recall} calculates the proportion of true duplicate matches in the candidate matching pairs $\{\mathcal{G}\}$ with respect to those in the groundtruth denoted $ \{\mathcal{G}^*\}$: $Recall = |\{\mathcal{G}\}\cap \{\mathcal{G}^*\}| /  \{\mathcal{G}^*\}|$.

\item \textbf{Precision} calculates the proportion of true matches in the candidate matching pairs $\{\mathcal{G}\}$ to the size of the candidate matching pairs: $Precision = |\{\mathcal{G}\}\cap \{\mathcal{G}^*\}| /  \{\mathcal{G}\}|$. 

\item \textbf{F1} is the harmonic mean between $Recall$ and $Precision$: $ {\rm F}1 = \frac{2 \times Recall \times Precision}{Recall + Precision}$. F1 score indicates the overall performance on both recall and precision.

\item \textbf{CSSR} is the candidate set size ratio which calculates the ratio of candidate set to the total number of possible candidates. We adapt the traditional CSSR formula in the relational setting 
by considering the nodes within a single graph. 
Our CSSR formula for graph data denoted $CSSR_g$ is defined as the ratio of the candidate set size to the number of possible candidate matches of the same node type from the graph \emph{i.e.} $ CSSR_g = \frac{|\{\mathcal{G}\}|}{(|\{V_l\}|\times|\{V_l-1\}|/2)} : V_l \subseteq V \land \forall v_i, v_j \in V_l, L(v_i) \asymp L(v_j)$. CSSR is often used to measure the efficiency of blocking \emph{i.e.} a smaller CSSR, \emph{ceteris paribus}, indicates that a smaller block captures the matches in the dataset, thus minimizing the amount of search required within a block.  

\item \textbf{Purity} is our proposed metric to measure the effectiveness of each block generated. For any given block $B$, \emph{purity} is the ratio of number of true duplicate matches to the size of the block, \emph{i.e} $Purity = \vert \{B\} \cap \{B^*\} \vert / \vert \{B\} \vert$, where $\{B\}$ denotes candidate pairs in block $B$, and $\{B^*\}$ represents the true matching pairs within block $B$. It is important to note that \emph{purity} is different from \emph{precision} in the sense that \emph{precision} is the overall ratio of true duplicate matches after all the blocks have been combined thus, \emph{purity} measures the ``in-block precision''.

\end{itemize}

\subsection{Effectiveness of GraphER}\label{sec:effectiveness}


\subsubsection{Performance on Graph ER Benchmark datasets}
\begin{figure}[!t]
    \vspace{-3ex}
    \centering
    \includegraphics[width=0.7\linewidth]{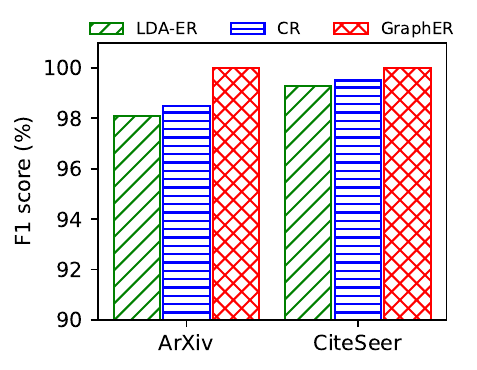}
    \vspace{-3ex}
    \caption{F1 Scores: ER Benchmark graph datasets}
    \label{fig:g-r}
    \vspace{-2ex}
\end{figure}

The F1 scores of the ER solutions over the benchmark graph datasets \emph{ArXiv} and \emph{CiteSeer} in 
Table~\ref{Tab:graphdatasets} are presented in Figure~\ref{fig:g-r}. In the figure we observe that {\sf GraphER} achieves better F1 scores (\emph{i.e.} 100\%) in comparison with \emph{LDA-ER} and \emph{CR} methods which are specifically designed for graphs. It is worth noting that due to the simplicity of \emph{ArXiv} and \emph{CiteSeer} datasets (\emph{i.e.} only one edge type), the recent deep learning based relational methods such as, 
\emph{DM}~\cite{1}, \emph{Ditto}~\cite{2}, and \emph{HG}~\cite{3} (not shown in Figure~\ref{fig:g-r}) can also achieve perfect F1 Scores. In subsequent sections we assess the performance of these techniques on more practical graphs with multiple edge types.




\begin{table}[t]
\caption{F1 Scores: complex graph datasets}
\label{tab:Effectiveness-graph}
\vspace{-2ex}
\resizebox{\linewidth}{!}{
\begin{threeparttable}
\begin{tabular}{@{}cccccccccc@{}}
\toprule
Dataset & DM & BERT & RoBERTa & Ditto & JointBERT & HG & RobEM & Certus & GraphER \\ \midrule
ER & $\sim$ & $\sim$ & \textbf{100} & 18.2 & $\sim$ & $\sim$ & 50.0 & 75.0 & \underline{88.9} \\
WWC-1 & 64.7 & 81.7 & 71.1 & 95.0 & 68.8 & \underline{96.2} & 71.9 & \textbf{100} & \textbf{100} \\
WWC-2 & \textbf{100} & 87.7 & 87.7 & \textbf{100} & 87.7 & \textbf{100} & \underline{92.1} & 79.8 & 84.9 \\
WWC-7 & 79.2 & 76.3 & 75.2 & \textbf{96.2} & 73.5 & 95.5 & 75.4 & 87.2 & \underline{95.6} \\
WWC-3 & 60.3 & 72.7 & 73.8 & \textbf{94.9} & 71.6 & \underline{93.7} & 73.8 & 85.2 & \underline{93.7} \\
WWC-6 & 32.3 & 74.3 & 55.2 & \underline{93.3} & 70.0 & 93.0 & 76.5 & 88.4 & \textbf{97.0} \\
WWC-5 & $\sim$ & 75.0 & 85.7 & 30.8 & $\sim$ & $\sim$ & \textbf{100} & 82.4 & \underline{91.9} \\
WWC-4 & $\sim$ & $\sim$ & 88.2 & 11.8 & $\sim$ & $\sim$ & 21.1 & \underline{88.9} & \textbf{100} \\
GDS-1 & \underline{95.6} & \textbf{100} & \textbf{100} & 95.0 & \textbf{100} & \underline{95.6} & \textbf{100} & \textbf{100} & \textbf{100} \\
GDS-2 & \textbf{100} & \textbf{100} & \textbf{100} & \textbf{100} & \textbf{100} & \textbf{100} & \underline{99.3} & 85.6 & 96.0 \\
GDS-7 & 91.8 & 88.3 & 88.0 & 96.5 & 87.7 & \textbf{98.2} & 90.0 & 87.2 & \underline{96.9} \\
GDS-3 & 86.4 & 89.1 & 89.8 & 94.0 & 83.3 & \underline{95.6} & 88.2 & 88.0 & \textbf{96.6} \\
GDS-6 & 73.9 & 80.6 & 73.5 & \underline{95.5} & 76.9 & 90.6 & 81.8 & 86.4 & \textbf{97.4} \\
GDS-5 & 72.7 & 66.7 & 72.7 & 72.7 & 66.7 & $\sim$ & 72.7 & \underline{90.6} & \textbf{95.5} \\
GDS-4 & $\sim$ & $\sim$ & 66.7 & 18.6 & $\sim$ & $\sim$ & 85.7 & \underline{86.7} & \textbf{97.0} \\ \midrule
Average & 57.1 & 66.2 & 81.9 & 74.2 & 59.1 & 63.9 & 78.6 & \underline{87.3} & \textbf{95.4} \\ \bottomrule
\end{tabular}
    \begin{tablenotes} 
		\item \emph{NB: Bold and underlined results indicate the best score and second best score respectively, and the label of $\sim$ means no result can be obtained.}
    \end{tablenotes}
\end{threeparttable}
}
\vspace{-2ex}
\end{table}

\subsubsection{Performance on Complex Graph ER datasets}

In this set of experiments, we assess the performance of the SOTA techniques (\emph{i.e.} \emph{DM}, \emph{BERT}, \emph{RoBERTa}, \emph{Ditto}, \emph{JointBERT}, \emph{HG}, \emph{RobEM}, and \emph{Certus}) in comparison to {\sf GraphER}
on the graph datasets in Table~\ref{Tab:graphdatasets}. 
The results are shown in Table~\ref{tab:Effectiveness-graph}. 

We observe that, in general, {\sf GraphER} performs significantly better (achieved an average F1 score of 95.4\%) than all the other SOTA techniques. This demonstrates the consistency of {\sf GraphER}’s performance across different scenarios exhibited by the
datasets. 
Considering the \emph{ER} dataset, \emph{RoBERTa} achieves the best F1 score of 100\%, and the second best of 88.9\% is achieved by {\sf GraphER}. However it is interesting to note that all other methods perform significantly poorer \emph{e.g.}  \emph{RobEM} and \emph{Ditto} only obtain 50\% and 18.2\% respectively, while 
\emph{DM}, \emph{BERT}, \emph{JointBERT} and \emph{HG} are not able to detect any of the duplicates.  This is because, the \emph{ER} dataset has only 12 nodes with duplicates representing only $\approx 1\%$ of the entire dataset. In this scenario, there is insufficient training samples for these techniques to learn from. On the other hand, {\sf GraphER} performs well indicating that it is able to handle such scenarios. 
This finding is further validated by considering \emph{WWC-7 (resp. GDS-7)}, \emph{WWC-3 (resp. GDS-3)}, \emph{WWC-6 (resp. GDS-6)}, \emph{WWC-5 (resp. GDS-5)}, and \emph{WWC-4 (resp. GDS-4)} corresponding to 20\%, 10\%, 5\%, 1\% and 0.5\% of nodes with duplicates respectively. 
One observes that as the proportion of duplicates decreases in the 
datasets, the performance of existing learning-based baseline techniques diminish. 
On the other hand, the rule-based method {\sf Certus} and our hybrid solution {\sf GraphER} are 
more robust as they require a relatively smaller amount of labelled data for rule synthesis, and
execution of their respective ER algorithms. 
It is worth noting that \emph{RoBERTa}, which is an optimised version of \emph{BERT}, seems to yield perfect scores on the \emph{ER} dataset, but as observed, this does not fully translate to other datasets.

Finally, to evaluate the impact of different types of noise on {\sf GraphER}, namely attribute and structural noise (\emph{cf.} Section~\ref{sec:datasets}), we compare the results of \emph{WWC-1 (resp. GDS-1)} with \emph{WWC-2 (resp. GDS-2)}. From the results, we can infer that {\sf GraphER} is least affected by attribute noise (\emph{i.e.}\emph{WWC-1 (resp. GDS-1)}) but more so with structural noise (\emph{i.e.}\emph{WWC-2 (resp. GDS-2)}).


\begin{table}[t]
\caption{F1 Scores: relational ER benchmark datasets}
\label{tab:Effectiveness-rela}
\vspace{-2ex}
\resizebox{\linewidth}{!}{
\begin{threeparttable}
\begin{tabular}{@{}cccccccccc@{}}
\toprule
Dataset & DM & BERT & RoBERTa & Ditto & JointBERT & HG & RobEM & Certus & GraphER \\ \midrule
FZ & \textbf{100} & \textbf{100} & \textbf{100} & \underline{98.1} & \textbf{100} & \textbf{100} & \textbf{100} & 88.6 & \textbf{100} \\
DA & 98.4 & 98.7 & 98.7 & \underline{99.0} & 98.1 & \textbf{99.1} & 98.5 & 97.8 & \underline{99.0} \\
AG & 69.3 & 71.0 & 70.8 & 74.1 & 70.4 & \underline{76.4} & \underline{76.4} & 65.8 & \textbf{80.5} \\
Beer & 72.7 & 84.8 & 84.8 & 84.6 & 78.8 & \textbf{93.3} & \underline{89.7} & 75.6 & 85.7 \\
DS & 94.7 & 94.5 & 95.7 & \underline{95.8} & 93.6 & \textbf{96.3} & \underline{95.8} & 91.9 & 92.3 \\
iA & 88.5 & 94.7 & 93.1 & 92.3 & 93.1 & \underline{96.3} & 96.1 & 95.2 & \textbf{97.1} \\
WA & 67.6 & 80.5 & 85.7 & 85.8 & 79.8 & \underline{88.2} & 85.5 & 80.7 & \textbf{88.6} \\ \midrule
Average & 84.5 & 89.2 & 89.8 & 90.0 & 87.7 & \textbf{92.8} & 91.7 & 85.1 & \underline{91.9} \\ \bottomrule
\end{tabular}
    \begin{tablenotes} 
		\item \emph{NB: Bold and underlined results indicate the best score and second best score respectively.}
    \end{tablenotes}
\end{threeparttable}
}
\vspace{-4ex}
\end{table}

\subsubsection{Performance on Benchmark Relational Datasets}


We show how the competing techniques performed over well-known relational ER datasets in Table~\ref{Tab:reladatasets}. The results of this evaluation is presented in Table~\ref{tab:Effectiveness-rela}. The average F1 scores (in \%) for each technique across all the datasets are 
shown within Table~\ref{tab:Effectiveness-rela}.

The average scores show that {\sf GraphER} achieves 91.9\% F1 across all the datasets, 
which is only second best to \emph{HG} (\emph{i.e.} 92.8\%). This demonstrates that {\sf GraphER} 
is not only a powerful tool for ER in graph data, but can also be used effectively in 
relational data. 
In this setting, {\sf Certus'} performance is comparable to 
\emph{DM} with the average F1 score of 84.5\%.
A deeper look at the results reveal yet another point of difference in 
{\sf GraphER}'s performance \emph{w.r.t.} all other techniques: it performs consistently 
well ($> 80\%$) over all the datasets. 
A head-to-head comparison of {\sf GraphER} and the best-performing model in this experiment, 
\emph{HG}, further highlight the prowess of our proposal (\emph{cf.} competition on Beer and AG datasets).

\subsection{Qualitative Analysis}
In this section, we aim to show the inherent explainability 
within {\sf GraphER} through a qualitative analysis of true positive (TP), true negative (TN), 
false positive (FP), and false negative (FN) results from four datasets, namely: \emph{Amazon-Google}, 
\emph{BeerAvo-RateBeer}, \emph{DBLP-Scholar}, and \emph{WWC-3}.

\subsubsection{Correct Matches: TP \& TN Matches}
\begin{figure*}[!t]
    \centering
    \includegraphics[width=1\linewidth]{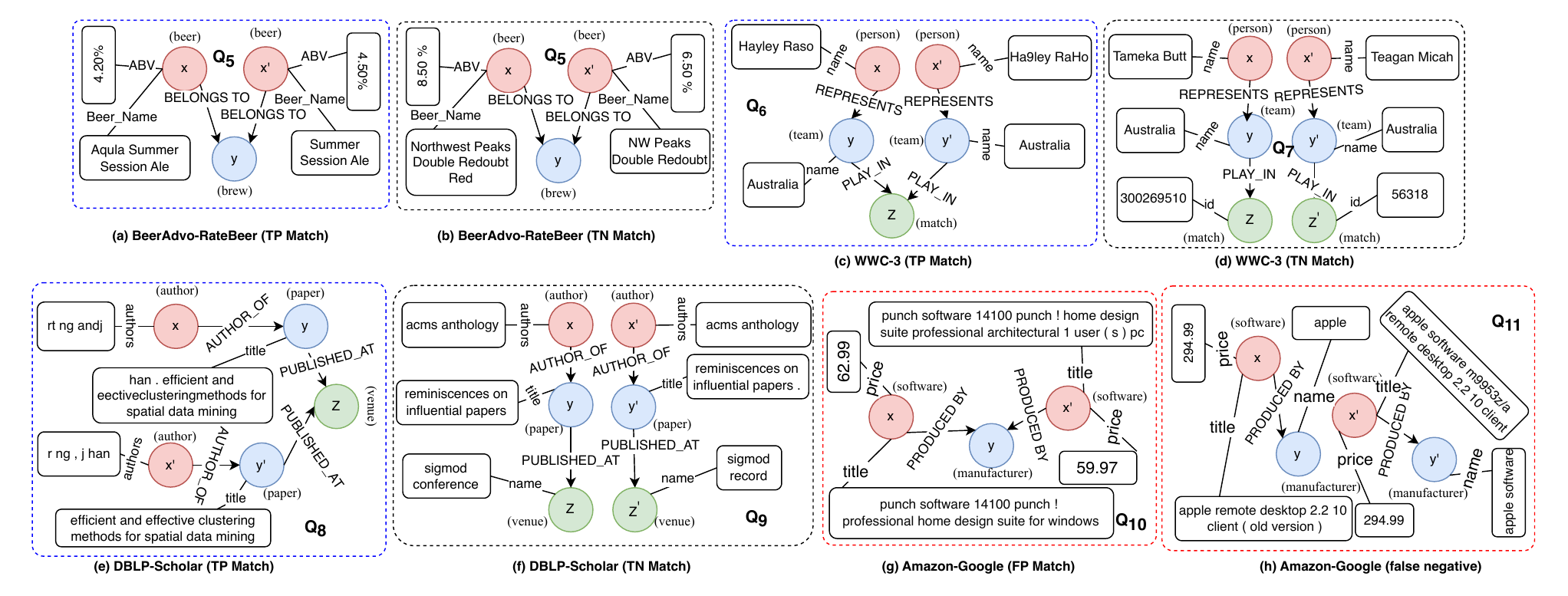}
    \vspace{-5ex}
    \caption{Exemplar matches for interpreting results}
    \label{fig:gp-intr}
    \vspace{-2ex}
\end{figure*}

%
Consider the GDD, $\varphi_3:Q_5[x,x',y], \delta_{BN}(x, 
\mbox{``Summer Session Ale''})\leq 0.3 \wedge \delta_{ABV}(x, x')\leq 1 
\to \delta_\e(x, x')=0$, over the \emph{BeerAvo-RateBeer} data. It specifies that for any two 
\texttt{beer} nodes, $x,x'$, with a \texttt{belong} relationship to a \texttt{brew} 
node $y$, i.e., a homomorphic match of $Q_5$: if the respective distances of 
$x,x'$ on the attribute \texttt{Beer\_Name} \emph{w.r.t.} \texttt{``Summer Session Ale''} as measured by $\delta_{BN}$ is within of $0.3$, 
and the difference in alcohol content of $x,x'$ is not more than $1$, then $x,x'$ 
are the same \texttt{beer}. 
A review of matches of pattern $Q_5$ in the Beer data with seemingly similar beer node 
attribute values show the efficacy of $\varphi_3$. For example, we show examples of 
TP and TN matches in Figure~\ref{fig:gp-intr}~(a) and (b) respectively. Whereas the
match in the former agrees with both the pattern and dependency of $\varphi_3$, the
latter violates the distance constraints in $\varphi_3$. Consequently, in both cases,
a correct prediction ensures: a true positive and a true negative match, respectively.

For the \emph{WWC-3} dataset, we employ the rule $\varphi_4: Q_6[x, x', y, y', z], \\
\delta_{N}(x, x')\leq 0.25 \wedge \delta_{N}(y, y') = 0 \to \delta_\e(x, x') = 0$ 
to explain the results shown in Figures~\ref{fig:gp-intr}~(c) and (d). $\varphi_4$
indicates that two \texttt{person} nodes, $x,x'$, with similar names who played for
two teams $y,y$ with exact names and played in the same \texttt{match} $z$ must refer
to the same real-world person. 
%
In the TP matching pair (Figure~\ref{fig:gp-intr}), the two players “Hayley Raso” and “Ha9ley RaHo” have slightly different names (possibly due to a typo) but do satisfy the distance constraint on name. The team names, “Australia”, are same, and played in the same match, and thus both names refer to the same real person, Hayley Raso. This example also illustrates the ability of {\sf GraphER} to point out the source of errors. In contrasts, the true negative pair shows two players with different names “Tameka Butt” and “Teagan Mitch” \emph{i.e.} they do not satisfy the distance constraint on name. Further, although they represent the same team “Australia”, they did not play in the same match \emph{i.e.} they do not satisfy the graph pattern $Q_6[x, x', y, y', z]$ providing further evidence of the correctness that the two names “Tameka Butt” and “Teagan Mitch” do not represent the same person.

In the \emph{DBLP-Scholar} graph, we use the rule $\varphi_5: Q_8[x, x', y, y', z], \\
\delta_{Au}(x, x')\leq 0.59 \wedge \delta_{Ti}(y, y') \leq 0.47 \to \delta_\e(y, y') = 0$. 
The rule states that for any match of $Q_8$ in the graph: if the \texttt{author} nodes $x,x'$ have similar list of author names, and the respective titles of the \texttt{paper}s 
$y,y'$ are also similar, then said papers $y,y'$ must be the same. 
This rule can be used to explain the matching results in Figure~\ref{fig:gp-intr}~(e) 
and (f). It is easy to see the example in Figure~\ref{fig:gp-intr}~(e) satisfies 
$\varphi_5$ , hence the \texttt{paper} nodes are linked. In fact, the authors and title of 
the two papers are within a distance of 0.59 and 0.47, respectively, and the papers are published in the same venue $z$. However, in the case of Figure~\ref{fig:gp-intr}~(f), 
although the distance constraints could be satisfied, the paper pairs are 
correctly determined as not same as the example does not support the structural
constraint, $Q_8$. 

\subsubsection{Incorrect Matches: FP \& FN Matches}
In the \emph{Amazon-Google} dataset, we use the GDD $\varphi_6 : (Q_9[x, x', y], \delta_{title}(x, x')\leq 0.54 \to  \delta_\e(x, x') = 0)$ to explain some incorrect
results. The rule specifies that two \texttt{software} nodes $x,x'$, \texttt{produced\_by}
the same \texttt{manufacturer}, $y$: $x,x'$ are the same piece of software 
if the dissimilarity of their titles is no more than 0.54. 
Thus, the node pair, $x,x'$, in Figure~\ref{fig:gp-intr}~(g) 
were incorrectly determined as the same since $\varphi_6$ is clearly satisfied. 
However, they are not actually a match in the \emph{Amazon-Google} dataset. 
This example illustrates the ``no-foolproof'' fail-cases of models, i.e.,
albeit rarely, some predictions (rule-aided/based or not) are wrong. 
Further, we use the FN case in Figure~\ref{fig:gp-intr}~(h) to highlight a key 
challenge for rule-based/-aided solutions. A careful review of the set, $\Sigma_{AG}$,
of mined rules in the \emph{Amazon-Google} dataset \emph{w.r.t.} the non-matching result for
the \texttt{software} nodes $x,x'$ in the diagram show that they did 
not constitute (or were embedded) in a match of any pattern over $\Sigma_{AG}$. 
Thus, $x,x'$ are simply labelled FN pairs as they did not satisfy any GDD.

\begin{figure*}[!t]
    \centering \subfigure{\includegraphics[height=3.5cm,width=4.2cm]{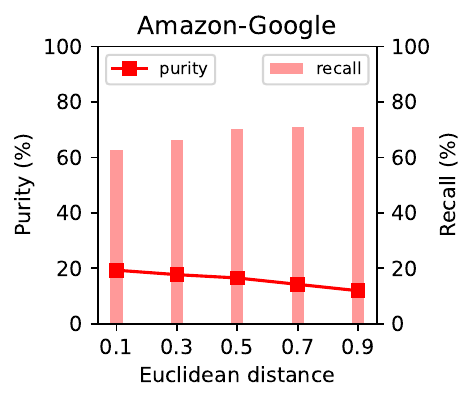}}
    \subfigure{\includegraphics[height=3.5cm,width=4.2cm]{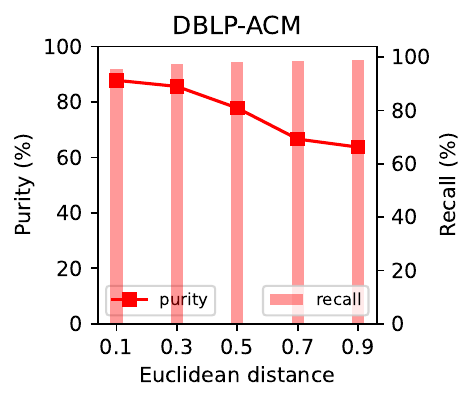}}
    \subfigure{\includegraphics[height=3.5cm,width=4.2cm]{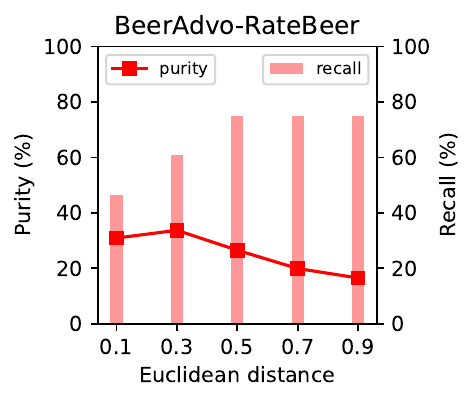}}
    \subfigure{\includegraphics[height=3.5cm,width=4.2cm]{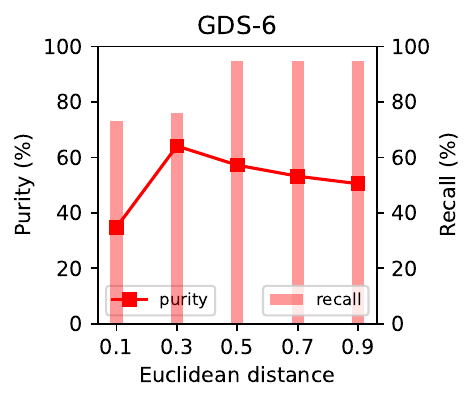}}
    \vspace{-3ex}
    \caption{Purity and recall in the blocking phase}
    \label{fig:blocking_phase}
    \vspace{-3ex}
\end{figure*}

\begin{figure*}[!t]
    \centering \subfigure{\includegraphics[height=3.5cm,width=4.2cm]{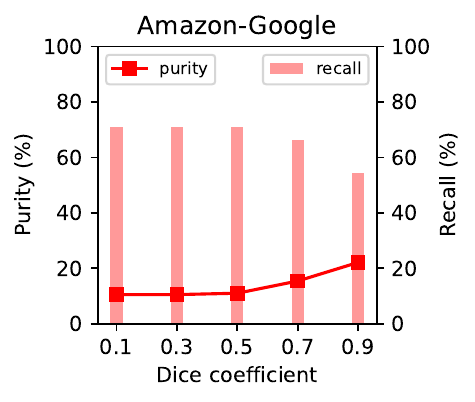}}
    \subfigure{\includegraphics[height=3.5cm,width=4.2cm]{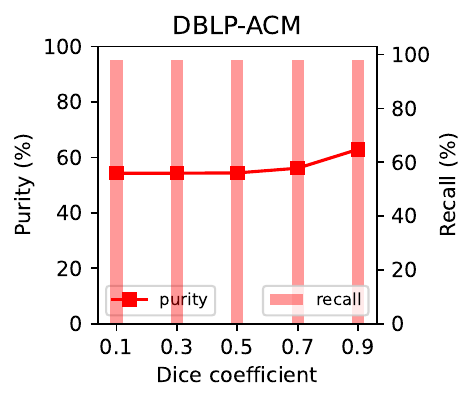}}
    \subfigure{\includegraphics[height=3.5cm,width=4.2cm]{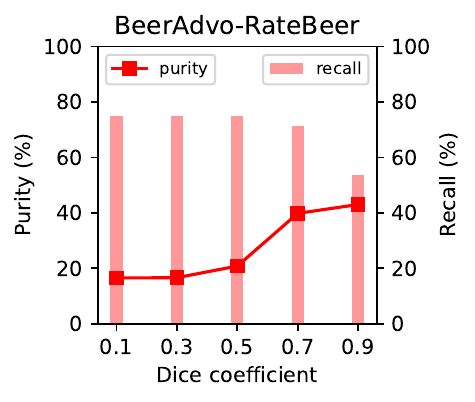}}
    \subfigure{\includegraphics[height=3.5cm,width=4.2cm]{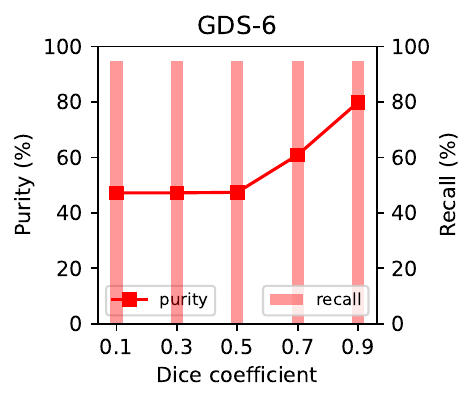}}
    \vspace{-3ex}
    \caption{Purity and recall in the pruning phase}
    \label{fig:pruning_phase}
    \vspace{-3ex}
\end{figure*}

\begin{figure*}[!t]
    \centering \subfigure{\includegraphics[height=3.5cm,width=4.2cm]{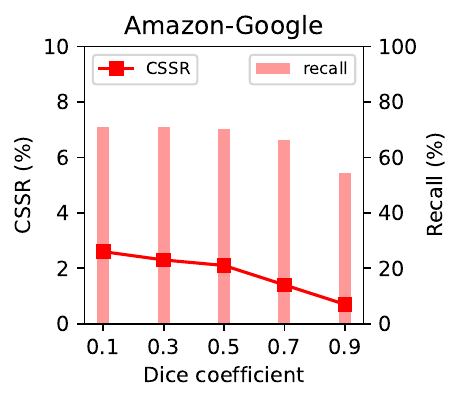}}
    \subfigure{\includegraphics[height=3.5cm,width=4.2cm]{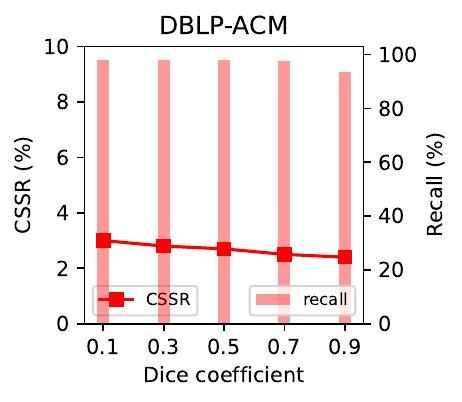}}
    \subfigure{\includegraphics[height=3.5cm,width=4.2cm]{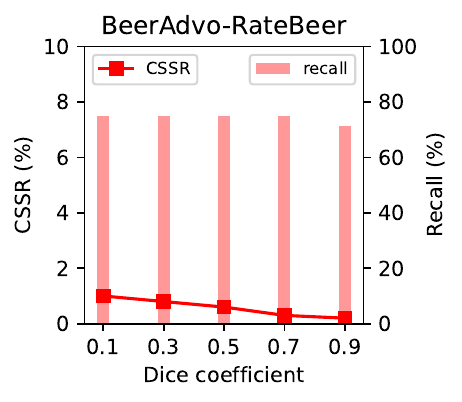}}
    \subfigure{\includegraphics[height=3.5cm,width=4.2cm]{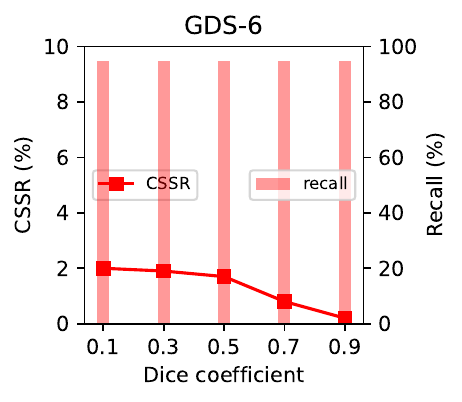}}
    \vspace{-3ex}
    \caption{CSSR and recall in the pruning phase}
    \label{fig:CSSR}
    \vspace{-3ex}
\end{figure*}

\begin{figure*}[!t]
    \centering
    \subfigure{\includegraphics[height=4cm,width=4cm]{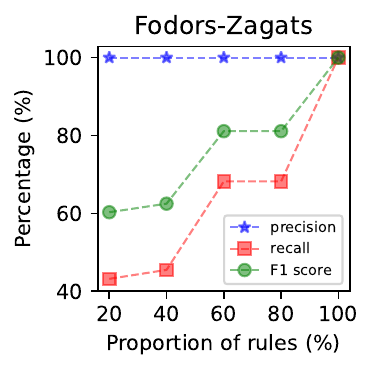}}
    \subfigure{\includegraphics[height=4cm,width=4cm]{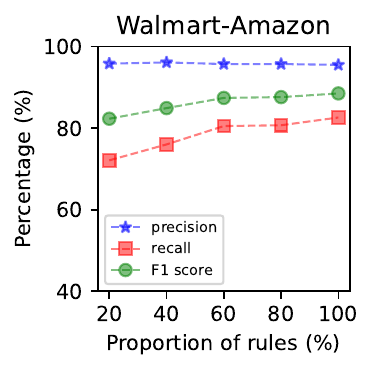}}
    \subfigure{\includegraphics[height=4cm,width=4cm]{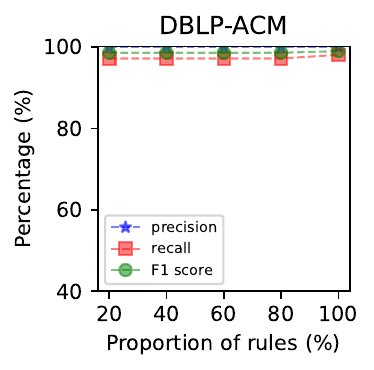}}
    \subfigure{\includegraphics[height=4cm,width=4cm]{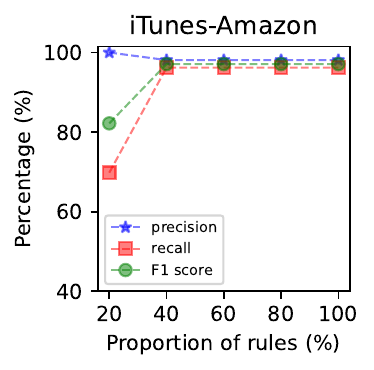}}
    \vspace{-3ex}
    \caption{Precision, recall, F1 score in the matching phase}
    \label{fig:matching_phase}
    \vspace{-2ex}
\end{figure*}

\subsection{Ablation Studies}\label{AS}
In this section, we assess the impact of the various phases of {\sf GraphER} on its effectiveness.

\subsubsection{Impact of Blocking, Pruning and Matching}
\textbf{Blocking}: 
To assess the impact of the blocking phase to the overall framework, we rely on our \emph{purity} metric. For each dataset, the purity of a block is controlled by adjusting the distance threshold parameter. The \emph{purity} for each block is calculated and then averaged, for all the blocks to give the overall purity score. Ideally, it is preferable to obtain high recall and high purity in the blocking phase.  Figure~\ref{fig:blocking_phase} is the results of this experiments. We notice that as the distance threshold is relaxed (increased), recall increases monotonically. Generally, as the threshold is relaxed, the purity of blocks decline, however it is possible that purity is improved (\emph{e.g.} \emph{GDS-6} and \emph{Beer}). This occurs when the additional vectors within a block as a result of the relaxation of the threshold are all duplicates.

\textbf{Pruning}: In the pruning stage, our goal is to remove mismatched entity pairs from all blocks while improving the purity of the blocks. Ideally, during the pruning phase, we want high purity while maintaining the same level of recall. Figure~\ref{fig:pruning_phase} shows the impact of the pruning threshold (dice coefficient) on purity and recall. As expected, the purity monotonically increases as the threshold is tightened. However this does not result in a severe loss in recall indicating the effectiveness of our pruning phase. Similarly, the CSSR score improves (reduces) when the threshold is tightened, however this does not significantly affect recall (\emph{cf.} Figure~\ref{fig:CSSR})

\textbf{Matching Phase:} As described in Section~\ref{sec:match}, in the matching phase, matches are confirmed by relying on the rules. We are interested to observe the changes to precision, recall, and F1 score as the percentage of matching rules from which a candidate pair must satisfy only one rule increases. Figure~\ref{fig:matching_phase} is the result of this experiment. The figure shows four subplots corresponding to four datasets, including two “clean” datasets (\emph{i.e.} Fodors-Zagats, DBLP-ACM) and two “dirty” datasets (\emph{i.e.} Walmart-Amazon and iTunes-Amazon). In the clean dataset, each node with duplicate matching node has only one match while in the dirty dataset, each node may have multiple duplicates. Each subplot is generated by determining a cutoff on the minimum proportion of rules of which one must be satisfied before a candidate matching pair is confirmed as a match. We observe that, as the number of rules increase, the recall generally increases. This is because the pool of rules from which at least one must be satisfied increases. However, the rate of improvement is different for each dataset. Interestingly, the precision is not severely affected. 
This indicates that, in most cases, by specifying only a few rules, {\sf GraphER} can achieve significant results.


\subsubsection{Impact of GDD-aided structural information and attribute information}
In this experiment, we aim to understand the contributions of both GDD-aided structural and attribute information to {\sf GraphER}. We develop two variants of {\sf GraphER} namely: (1) \emph{GraphER-S}, where only the GDD-aided structural information (\emph{cf.} component \encircle{A}) in the form of graph patterns. That is, the attributes auto-encoder is excluded 
from the vectorization stage;  (2) \emph{GraphER-A} where only GDD-aided attribute information (\emph{cf.} component \encircle{B}) is used. Thus, in the vectorization stage, the meta-path-based skip-gram is excluded. In both cases, the confirmation of matches using rules (\emph{cf.} component \encircle{C}) are excluded. A $k$-nearest neighbor approach based on cosine measure is used in its stead, where $k$ depends on the number of expected matches in the ground truth. 
Table~\ref{Tab:variant} is the result of the experiments and the bold result indicates the best performance. 
From the Table, we notice that by combining both GDD-aided structural and attribute information, there is considerable improvement (\emph{i.e.} {\sf GraphER}) compared to using only one (\emph{i.e.} \emph{GraphER-S} or \emph{GraphER-A} alone).

\begin{table}[ht]
\caption{F1 scores: different variants on datasets}
\vspace{-1.5ex}
\label{Tab:variant}
\resizebox{\linewidth}{!}{
\begin{tabular}{@{}l|cccccccccc@{}}
\toprule
 & FZ & DA & AG & Beer & DS & iA & WA & ER & WWC-6 & GDS-6 \\ \midrule
GraphER-S & 85.5 & 95.8 & 56.3 & 63.4 & 60.9 & 75.3 & 58.5 & 80.0 & 92.3 & 72.3 \\
GraphER-A & 85.0 & 96.2 & 65.3 & 75.6 & 70.6 & 85.3 & 67.2 & 60.0 & 83.5 & 68.9 \\
GraphER & \textbf{100} & \textbf{99.0} & \textbf{80.5} & \textbf{85.7} & \textbf{92.3} & \textbf{97.1} & \textbf{88.6} & \textbf{88.9} & \textbf{97.0} & \textbf{97.4} \\ \bottomrule
\end{tabular}
}
\vspace{-2ex}
\end{table}

\subsection{Efficiency}

In this experiment we aim to understand the efficiency of {\sf GraphER} in \emph{w.r.t.} other techniques. We recorded the time (in seconds) for each algorithm across all datasets. This is shown in Table~\ref{Tab:efficiency} (end-to-end ER time performances). 
From the Table,  it is obvious that \emph{DM}, \emph{BERT}, \emph{JointBERT}, and \emph{HG} have no time performances on the \emph{ER} dataset due to insufficient training data (unsuccessful execution). We notice that on average {\sf GraphER} is as efficient as \emph{DM} and \emph{RobEM}, and is approximately six times less efficient than the highest efficiency \emph{Ditto}. 


\begin{table}[!t]
\caption{Run-time experiments}
\vspace{-1.5ex}
\label{Tab:efficiency}
\resizebox{\linewidth}{!}{
\begin{threeparttable}
\begin{tabular}{@{}cccccccccc@{}}
\toprule
Dataset & DM & BERT & RoBERTa & Ditto & JointBERT & HG & RobEM & Certus & GraphER \\ \midrule
FZ & 607s & 625s & 1079s & \textbf{81s} & 677s & 223s & 442s & 612s & 760s \\
DA & 9305s & \textbf{764s} & 1448s & 1248s & 1424s & 3222s & 6750s & 2771s & 5512s \\
AG & 3399s & 748s & 857s & 915s & \textbf{841s} & 1378s & 6113s & 1412s & 4525s \\
Beer & 227s & 762s & 217s & \textbf{64s} & 319s & 128s & 242s & 1856s & 2122s \\
DS & 16240s & \textbf{1500s} & 1627s & 2754s & 1975s & 6210s & 20412s & 10170s & 15950s \\
iA & 690s & 258s & 389s & \textbf{71s} & 619s & 264s & 312s & 2664s & 3383s \\
WA & 5193s & \textbf{856s} & 901s & 1005s & 1046s & 3392s & 5348s & 1408s & 2406s \\
ER & $\sim$ & $\sim$ & 894s & \textbf{48s} & $\sim$ & $\sim$ & 89s & 524s & 572s \\
WWC-6 & 373s & 381s & 676s & \textbf{141s} & 636s & 164s & 447s & 2308s & 2701s \\
GDS-6 & 1135s & \textbf{193s} & 543s & 295s & 215s & 429s & 806s & 2779s & 4870s \\ \midrule
Average & 4130s & 676s & 863s & \textbf{662s} & 861s & 1712s & 4096s & 2650s & 4280s \\ \bottomrule
\end{tabular}
\begin{tablenotes} 
		\item \emph{NB: Bold result indicates the best time performance, and the label of $\sim$ means no result can be obtained.}
    \end{tablenotes}
\end{threeparttable}
}
\vspace{-3.5ex}
\end{table}

\section{Related Work}\label{sec:rel}
This paper sits at the intersection of three vibrant fields: graph constraints learning (GCL), graph representation learning (GRL), and 
entity linking and resolution (ELR). 

\subsubsection*{\bf GCL}
The proposed ELR framework in this work hinges on the existence of semantically 
meaningful graph rules. Indeed, graph constraints have been the subject of 
research in recent years: Graph keys (GKeys)~\cite{GKeys} are a 
class of keys for graphs based on isomorphic graph properties for identifying unique entities in graphs; Graph functional dependencies (GFDs)~\cite{GFDs} impose attribute-value dependencies (like conditional functional dependencies (CFDs)~\cite{CFDs}) upon topological structures in graphs; and Graph entity dependencies (GEDs)~\cite{9} unify and subsume the semantics of both GFDs and GKeys. 
Other rules such as graph quality rules (GQRs)~\cite{GQRs} extend GEDs by supporting inequality literals, to 
reduce false positives in object identification \emph{i.e.}  entities that do not match; Graph-pattern association rules (GPARs)~\cite{GPARs} extend the association rules of itemsets to discover social graphs and identify potential customer social impacts through exploration; and Graph association rules (GARs)~\cite{GARs, Big-GARs} extend GPARs with preconditions and GFDs with limited existential semantics. 
Clearly, there is a growing list of graph languages, and their discovery techniques
(e.g.,~\cite{10,zhang2023discovering,FASTAGEDs}) that can be used to encode various 
data semantics within the proposed ELR framework.
In this work, we leverage GDDs~\cite{6} which extend GEDs to include the semantics 
of similarity and matching for use as declarative matching rules.

\subsubsection*{\bf GRL} 
GRL techniques have been shown to be effective in many tasks, such as classification, link prediction, and matching~\cite{GL-survey1, GL-survey2, GL-survey3, GL-survey4}. Generally, graph learning methods use machine learning algorithms to extract relevant features of graphs. There are three approaches: a) matrix factorization based methods such as~\cite{GF, Laplacian, GraRep, HOPE} which adopt a matrix to represent graph characteristics like vertex and pairwise similarity, and embeddings generated by factorizing this matrix~\cite{he2003locality}; b) random walk based methods which has been shown to be a convenient and effective way to sample networks~\cite{xia2014mvcwalker, xia2019random}, includes works such as~\cite{metapath2vec, hin2vec, deepwalk, node2vec, LINE, struc2vec}. These generate node sequences while maintaining the original relationship between nodes, and then generate feature vectors of vertices, so that downstream tasks can mine network information in low-dimensional space; and c) message passing based methods are a general framework for graph neural networks~\cite{GNN}, which follows the "message passing paradigm", that is, nodes update their own feature vectors by exchanging information with their neighbors. This paradigm can include many variants of graph neural networks, such as graph auto-encoders (GAE~\cite{GAE}), graph convolutional networks (GCN~\cite{GCN, GraphSAGE}), graph attention networks (GAT~\cite{GAT}), etc. In this work, we employ knowledge-driven random walk and message passing to implement structural embedding and attribute embedding for property graphs. 

\subsubsection*{\bf ELR} 
The ELR problem has been widely studied within the relational data setting 
(\emph{e.g.}~\cite{1, 2, Auto-em}), with growing attention in the graph data 
setting (\emph{e.g.}~\cite{5,6}). In general, the ER task is often studied under 
two main sub-task: blocking and matching.
The aim of blocking is to generate rough groupings of possible candidate pairs
to ensure efficient evaluation. Predominate blocking 
techniques include: (deep) learning-based~\cite{7,21,22,23}, and various heuristic 
rules~\cite{18,19,20}.
On the other hand, the goal of the matching problem is to determine whether or not
a given pair of candidates is a true match. Similarly, existing ELR matching works 
can also be categorise as learning~\cite{1,2,3,4} or non-learning~\cite{24,25,26} approaches. 
In both (blocking and matching) cases, the latter methods are usually easier to 
understand than the former in many practical scenarios. However, the learning
based methods often provide the best accuracy, albeit their explainability and/or
interpretability are non-trivial and require further models like~\cite{ebaid2019explainer, teofili2022effective}. 
A few works use extraneous knowledge bases to augment the linking process~\cite{EL}, 
including the works in~\cite{ganea2016probabilistic, guo2014entity, fang2019joint}. 
Others have focused on identifying equivalent entities amongst different knowledge graphs~\cite{zeng2021comprehensive} including the representation learning-based alignment methods~\cite{lin2019guiding, liu2022selfkg, chen2016multilingual, wang2018cross}. 
Also noteworthy is the collective ELR setting~\cite{cao2018neural, bhattacharya2007collective}, 
in line with Dedupalog~\cite{arasu2009large}, the work in~\cite{bienvenu2022lace} proposes a declarative logical framework, LACE which can employ hard and soft rules to determine when 
pairs of entities should be matched.

The most related ELR works to this work are those in~\cite{6,15}. 
In~\cite{6}, graph differential dependencies (GDDs) are learnt and used to resolve both 
graph and relational entities; whereas our proposed solution, {\sf GraphER}, is a novel 
framework for using both graph rules- (e.g., GDDs) and graph representation learning- 
(e.g., GNNs) techniques to resolve entities. This enables {\sf GraphER} to achieve 
levels of accuracy comparable to learning-based methods, with an added benefit of intuitive
interpretability and explainability of rule-based methods. On the other hand,
the work in~\cite{15} seeks to achieve the same goal as {\sf GraphER} by encoding 
machine learning ELR models as the explicit consequence of logic rules. 
They only propose a hybrid framework that can be used, but no actual method is proposed and no experiment is conducted.

Meanwhile, ~\cite{christophides2015entity} and ~\cite{christophides2020overview} emphasize ER on RDF graphs which poses a different set of challenges compared to property graphs e.g., whereas RDF graphs embed a triple structure with each triple identified by a URI, this is not the case in property graphs. ~\cite{kirielle2023unsupervised} do not solve the problem of ER in property graphs but rather use a graph-model of dependencies as a tool to solve ER challenges in relational data.

\section{Conclusion}\label{sec:con} 
We propose an effective ER solution for graph data, called {\sf GraphER}. 
Our solution derives synergy from graph constraints (GDDs) and graph 
neural representation learning techniques (GNNs) to enhance both the effectiveness and explainability of ER in property graphs. 
Our comprehensive experiments show that not only is {\sf GraphER} superior in graph data, but it also achieves highly competitive results in relational data. 
As the volume of unstructured data increases disproportionately to structured data, it is expected that ER techniques such as {\sf GraphER} that are more amenable to convenient forms of data representation, \emph{i.e.} graphs, will play an important role in ER tasks. In the future, we will investigate how large language models can be used to enhance the ability of GNNs, while simultaneously learning the structure and attribute information of property graph.

\vspace{-2pt}


\balance

\bibliographystyle{ACM-Reference-Format}
\bibliography{reference}
\end{document}